\documentclass{PoS}

\title{Standard Model Heavy Flavor physics on the Lattice}

\ShortTitle{Standard Model Flavor physics on the Lattice}

\author{\speaker{Christine Davies}%
\\
University of Glasgow\\
E-mail: \email{c.davies@physics.gla.ac.uk}}

\abstract{ Lattice QCD calculations in charm and bottom physics are 
particularly important because they can provide the hadronic 
weak decay matrix elements needed for key constraints on the CKM Unitarity Triangle. 
I will summarise recent results in this area,
paying particular attention to 
sources of error, comparison between methods and tests of results 
against experiment, for example, in the spectrum. 
Updated world averages for decay constants 
this year are :  $f_{D_s}$=248.6(2.4) MeV; $f_D$ = 212.1(3.4) MeV; 
$f_{B_s}$ = 227(4) MeV; $f_B$ = 190(4) MeV. 
Note that $B$ decay constants are clearly lower than the 
corresponding $D$ decay constants. 
Improved $D$ semileptonic 
form factors, both shape and normalisation, now allow the 
direct determination of $V_{cs}$ and $V_{cd}$ to 3\% and 5\% respectively. 
This year we also have a clear demonstration 
that dependence of form factors on the spectator quark mass between 
light and strange is very small. Apart from the phenomenology implications, 
this has practical application to the normalisation of branching 
fractions in experiment.
The current best Standard model rate for a 
key LHC mode sensitive to new physics - 
$B_{(s)} \rightarrow \mu^+\mu^-$ - is derived from  
lattice QCD calculations on $B/B_s$ mixing rates. I will discuss the 
current result and prospects
for improving lattice QCD errors. 
}

\FullConference{XXIX International Symposium on Lattice Field Theory \\
		 July 10 -- 16 2011\\
		Squaw Valley, Lake Tahoe, California}

\begin{document}

\section{Introduction}

Heavy quark physics has turned out to be one of the `killer 
applications' of lattice QCD and results in this area have done 
much to persuade the particle physics community that we 
now have a serious tool for calculating strong interaction 
effects that can provide accurate phenomenology not available 
with any other method. Particularly important are 
various heavy meson weak decay matrix elements that are 
key to constraining the vertex of the Unitarity Triangle 
derived from the Cabibbo-Kobayashi-Maskawa (CKM) matrix for a 
stringent test of the self-consistency of the Standard Model. 
Calculations of these matrix elements must not be seen in isolation, however. 
One of the key features of (lattice) QCD is the 
small number of free parameters - a mass for each quark 
and an overall scale parameter or equivalently a coupling constant. 
Once these are fixed a myriad of parameter-free tests against experiment 
become possible. This is particularly true in the heavy quark 
sector where there are many gold-plated states in the spectrum. 
It is still possible to make predictions ahead of experiment, 
which are many times more valuable than postdictions 
in terms of credibility. In addition new methods for heavy 
quarks have allowed us to leverage improved accuracy in light 
quark physics, for example for quark masses.  

Here I will discuss the current status of lattice QCD calculations 
in charm and bottom physics, comparing formalisms, providing world 
averages and discussing prospects for the future. An important 
theme will be how to improve the errors {\it and} how to test that 
we have improved the errors, since precision from lattice QCD is critical for 
the Unitarity Triangle tests discussed above.  

\section{Heavy quark physics on the lattice}

In discretising the QCD Lagrangian onto a space-time lattice 
we inevitably generate discretisation errors that appear 
as some power of the lattice spacing, $a$. 
Physical results, for example hadron masses, then
depends on the lattice spacing as:
\begin{equation}
m(a) = m_{a=0}\left[ 1 + A(\Lambda a)^i + B(\Lambda a)^j + \ldots \right] .
\label{eq:disc}
\end{equation}
Here, for a light hadron, we would expect $\Lambda$ to be the 
typical dimensionful scale of QCD, say a few hundred MeV. 
Then $A, B$ are $\mathcal{O}(1)$. 
Since the lattice spacing of a modern lattice QCD calculation 
is of size $a^{-1} \approx 1 - 3$ GeV, $\Lambda a << 1$. 
Good discretisations, with small errors, have leading 
power $i=2$ and higher powers, $j$, 
starting at 4. 

For heavy quarks the scale for the discretisation errors will 
typically be set by the heavy quark mass, $m_Q$, which makes 
controlling the discretisation errors harder. For a 
lattice spacing $a \approx 0.1 {\rm fm}$ then $m_ca \approx 0.4$, 
$m_ba \approx 2$. For charm quarks this indicates that, although 
discretisation errors will be larger than those for light hadrons, 
good results are possible with a highly improved action on 
fine lattices. For example, using the Highly Improved Staggered 
Quark action~\cite{hisq} where 
the leading errors are $\mathcal{O}(\alpha_sa^2)$, 
discretisation errors of 2\% are seen on 0.09fm lattices 
(where $m_ca = 0.4$) in the decay constant of 
the $\eta_c$ meson, when using the heavy quark potential 
parameter, $r_1$, to fix the lattice spacing~\cite{fds}. For light meson 
decay constants at the same lattice spacings, 
the discretisation errors are barely visible~\cite{r1}.  
For $b$ quarks the situation is worse and 
this is why methods using a nonrelativistic expansion 
of the Dirac Lagrangian were developed. 
By removing $m_Q$ as a dynamical scale in the calculation 
they allow the scale of discretisation errors to be set again 
by $\Lambda$. A price has to be paid in terms of complexity 
and in other sources of  
systematic error coming from the nonrelativistic expansion. 
However, these methods have been the workhorses of $b$ physics 
in the past because they enabled calculations to be done at 
the lattice spacing value that were available. As we will see here, this is 
is beginning to change 
but it is likely that we will continue to 
need a mix of methods into the future. 

There are a variety of relativistic actions 
in use for light quarks and charm quark physics programmes 
are now being developed with all of them, sometimes with 
some modifications. 
\begin{itemize}
\item{The Highly Improved Staggered Quark action developed 
by the HPQCD collaboration~\cite{hisq} includes a further smearing level 
beyond the asqtad improved staggered quark action to give 
discretisation errors at $\mathcal{O}(\alpha_sa^2)$ and $\mathcal{O}(a^4)$, 
with small taste-changing errors. When used for heavy quarks, 
the coefficient of the `Naik' 3-link improvement term is 
modified to have a coefficient which is calculated as an expansion 
in $ma$ (starting at $(ma)^2$) to remove the leading $(ma)^4$ 
errors. Results from extending this action to $b$ quarks are now available~\cite{curr, fbs}. }
\item{The twisted mass action developed by the European Twisted 
Mass Collaboration~\cite{tm} uses a doublet of Wilson-like quarks that have an 
additional mass term multiplying $\tau_3$ in flavor space. 
This is used to introduce a $u/d$ doublet in the sea and work 
is underway on configurations that also include a $s/c$ doublet~\cite{tmnf4}. 
For valence $c$ and $s$ quarks ETMC use a separate doublet for 
each flavor~\cite{tmcnf2}.
The leading discretisation errors are ${\mathcal{O}}(a^2)$ at 
maximal twist~\cite{tm}. This action is also being used now 
for heavier quarks to extrapolate up to $b$~\cite{tmfb}. }
\item{Clover actions use a clover term to remove the tree-level $\mathcal{O}(a)$ 
errors from the Wilson action and have been in use for many years. In 
principle, unless the clover term is tuned nonperturbatively, these 
actions have $\alpha_sa$ discretisation errors. In newer 
variants, including various kinds of smeared links, 
discretisation errors can be made quite small and effectively 
$\mathcal{O}(a^2)$ and $\mathcal{O}(a^3)$~\cite{durr}. The clover action is 
being used for $c$ quarks on lattices with a fine temporal lattice 
spacing and relatively coarse spatial lattice spacing by the Hadron 
Spectrum collaboration~\cite{ryan}. }
\item{The good chiral properties of domain wall and overlap quarks enforce 
discretisation errors starting at $\mathcal{O}(a^2)$. This 
is an expensive method for charm physics, but is being tested. 
See, for example~\cite{mathur}.  }
\end{itemize}

The advantages of a relativistic action for $c$ quarks, if 
discretisation errors can be made small, are:
\begin{itemize}
\item{The hadron mass is simply and precisely obtained from the energy of 
the zero momentum hadron correlator. With a non-relativistic approach it 
is necessary to calculate finite-momentum correlators and extract 
the `kinetic mass' from the momentum dependence of the energy which is 
much less precise. } 
\item{In formalisms with enough chiral symmetry (HISQ, twisted mass 
and domain wall/overlap in the list above) the PCAC relation 
protects the axial current from renormalisation.  This means, for example, that 
pseudoscalar decay constants can be obtained directly with the correct 
normalisation and no error from a $Z$ factor is required.}
\item{Using the same action for $c$ quarks as for $u/d$ and $s$ allows 
some cancellation in ratios. For example the HPQCD collaboration obtained 
$m_c/m_s$ to 1\% (value 11.85(16)) using the HISQ action 
for both quarks~\cite{mcs}. ETMC obtain 12.0(3)~\cite{tmcnf2} for this ratio 
and 11.34(45) is a preliminary result with `Brillouin-improved' 
Wilson quarks~\cite{koutsou}.} 
\end{itemize}

The contrasting approaches that incorporate nonrelativistic ideas are:
\begin{itemize}
\item{NonRelativistic QCD (NRQCD) is a discretised version of a 
nonrelativistic effective theory which can be matched at a given order 
in $v_h$, the velocity of the heavy quark, to full QCD~\cite{nakhleh}. The heavy quark 
mass is tuned nonperturbatively and this fixes the $\mathcal{O}(v_h^2)$ 
kinetic energy term. Higher order terms have coefficients that are 
in principle calculable in perturbation theory (and will typically 
diverge as $m_ha \rightarrow 0$) but in most work 
to date have taken tree level values. The same action can be used 
for heavy-heavy and heavy-light physics. } 
\item{The discretisation of Heavy Quark Effective Theory onto the lattice 
starts from the static (infinite mass) approximation and adds $1/m_h$ and 
higher corrections through the calculation of matrix elements, with 
coefficients determined nonperturbatively, rather 
than incorporating kinetic terms in the dynamics. It can be used for heavy-light 
physics where a systematic programme has been developed by the 
Alpha collaboration~\cite{alpha}. }
\item{The Fermilab method, as originally implemented, 
uses the tadpole-improved clover action with a heavy quark 
mass ($b$ or $c$) but 
removing the leading discretisation errors by fixing the quark mass 
from the meson kinetic energy~\cite{fermilab}. The field is also 
`rotated' to remove tree-level $\mathcal{O}(a)$ errors. As the lattice 
spacing is reduced this becomes the standard clover action. Extensions 
of this method called `Relativistic Heavy Quarks' (RHQ) tune further 
coefficients nonperturbatively~\cite{rhqjap}, 
for example fixing the clover term 
from a hyperfine splitting and an asymmetry between time and space 
so that static and kinetic masses are equal~\cite{rhqrbc}.} 
\end{itemize}
It can be helpful for an accurate comparison of $b$ and $c$ physics to 
use the same action for both. 
However, so far this has really only been made an advantage 
in results using the relativistic HISQ action for all 5 quarks~\cite{curr}.  

\begin{figure}[h]
\parbox{0.5\hsize}{
$$a) \left( \begin{array}{ccc} V_{ud} & V_{us} & V_{ub} \\ \pi \rightarrow l \nu & K \rightarrow l \nu & B \rightarrow l \nu \\   & K \rightarrow \pi l \nu & B \rightarrow \pi l \nu \\ V_{cd} & V_{cs} & V_{cb} \\ D \rightarrow l \nu & D_s \rightarrow l \nu &  B_c \rightarrow l \nu \\ D \rightarrow \pi l \nu & D \rightarrow K l \nu & B \rightarrow D l \nu \\ V_{td} & V_{ts} & V_{tb} \\ \langle B_d | \overline{B}_d \rangle & \langle B_s | \overline{B}_s \rangle &  \\ \end{array} \right) $$
}
\hspace{10mm}
\parbox{0.5\hsize}{
$b)$ \includegraphics[width=0.6\hsize]{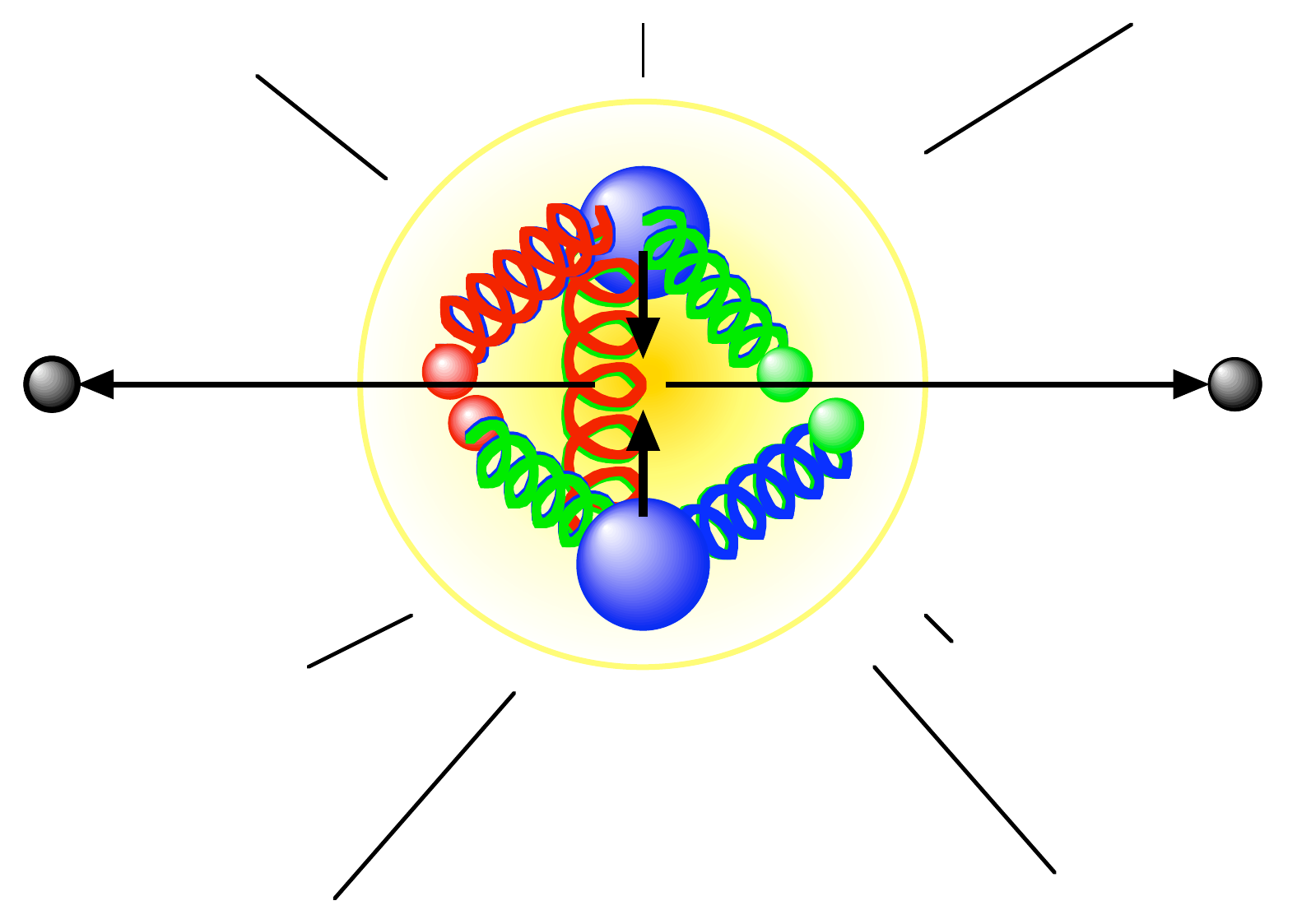}
}
\caption{ a) The Cabibbo-Kobayashi-Maskawa matrix in the Standard Model and the simple 
weak decay processes from which the elements can be determined given accurate lattice QCD calculations. b) an illustration of the leptonic decay of a charged pseudoscalar meson 
through annihilation of its valence quark and antiquark to a $W$ boson.  }
\label{fig:ckm}
\end{figure}

A key aim of flavor physics from lattice QCD is to calculate simple 
meson weak matrix elements that allow the determination of elements of 
the CKM matrix from experimental rates for 
leptonic or semileptonic decays or (for neutral $B/K$ mesons) oscillations. 
In Fig.~\ref{fig:ckm} I give the CKM matrix and the simple processes that allow the 
determination of that CKM element by combining experiment and lattice QCD. 
The processes are dominated by those for $B$ and $D$ meson decay, hence the 
importance of $b$ and $c$ quark physics in lattice QCD. Here I will review 
lattice calculations for a number of these processes 
and summarise the direct tests of CKM unitarity that result.  
Further discussion of the impact of these tests on Beyond the Standard 
Model scenarios is given in Enrico 
Lunghi's talk~\cite{lunghi}. 

\section{Charm physics results}

\subsection{Spectroscopy}
The spectroscopy of charmonium, $D$ and $D_s$ mesons and charmed 
baryons has had a resurgence of interest from experiment in 
recent years, particularly following the discovery of a number 
of new states in the charmonium spectrum by the B factories. 
The LHC will produce charmed and doubly charmed baryons in huge 
numbers, giving lattice QCD the opportunity to predict some 
masses ahead of experiment. 

This year saw new general spectroscopy results from a 
variety of clover actions. The clover action, as discussed above, 
is not highly improved and so discretisation errors are not 
minimised, but it is a simple and relatively efficient action to use when calculating 
a lot of hadron correlators with multiple source and sink 
operators. Examples include promising first results from the 
Hadron Spectrum Collaboration~\cite{ryan} giving 
a large number of states in the charmonium spectrum. 
The use of an anisotropic lattice with a finer spacing in the time 
than in spatial directions enables improved access to excited states, 
which can otherwise disappear rapidly from the correlator as a function 
of time from the source. In addition many operators are used for 
each irrep of the lattice rotation group, allowing identification of 
the expected continuum multiplets and the separation of 
quark model and hybrid states. 
All of this requires an enormous number of correlators to be calculated, 
however, and so results at only one lattice spacing and sea $u/d$ quark 
mass (on $n_f=2+1$ configurations) are available so far. 
The issue of overlap with multi-hadron channels for the 
excited `non-gold-plated' states has also not yet been 
addressed. The X(3872)~\cite{x3872} is still safe, but perhaps not for much 
longer, from the 
straitjacket of an unambiguous lattice QCD identification. 

The aims of calculations using SLiNC and 2-HEX smeared clover 
quarks on isotropic $n_f=2+1$ configurations 
by~\cite{rubio} are similar 
but also include preliminary results on 
the $D$ and $D_s$ spectrum. Results for charmed baryons 
were given by~\cite{briceno} using an RHQ 
action on the new `second generation' 
MILC configurations that include 2+1+1 flavors of sea HISQ quarks.  
The conclusions back up earlier indications that 
the mass of the $\Xi_{cc}$ found by SELEX~\cite{selex} is 
unlikely to be correct. 

\begin{figure}[h]
\parbox{0.5\hsize}{
$a)$ \includegraphics[width=0.9\hsize]{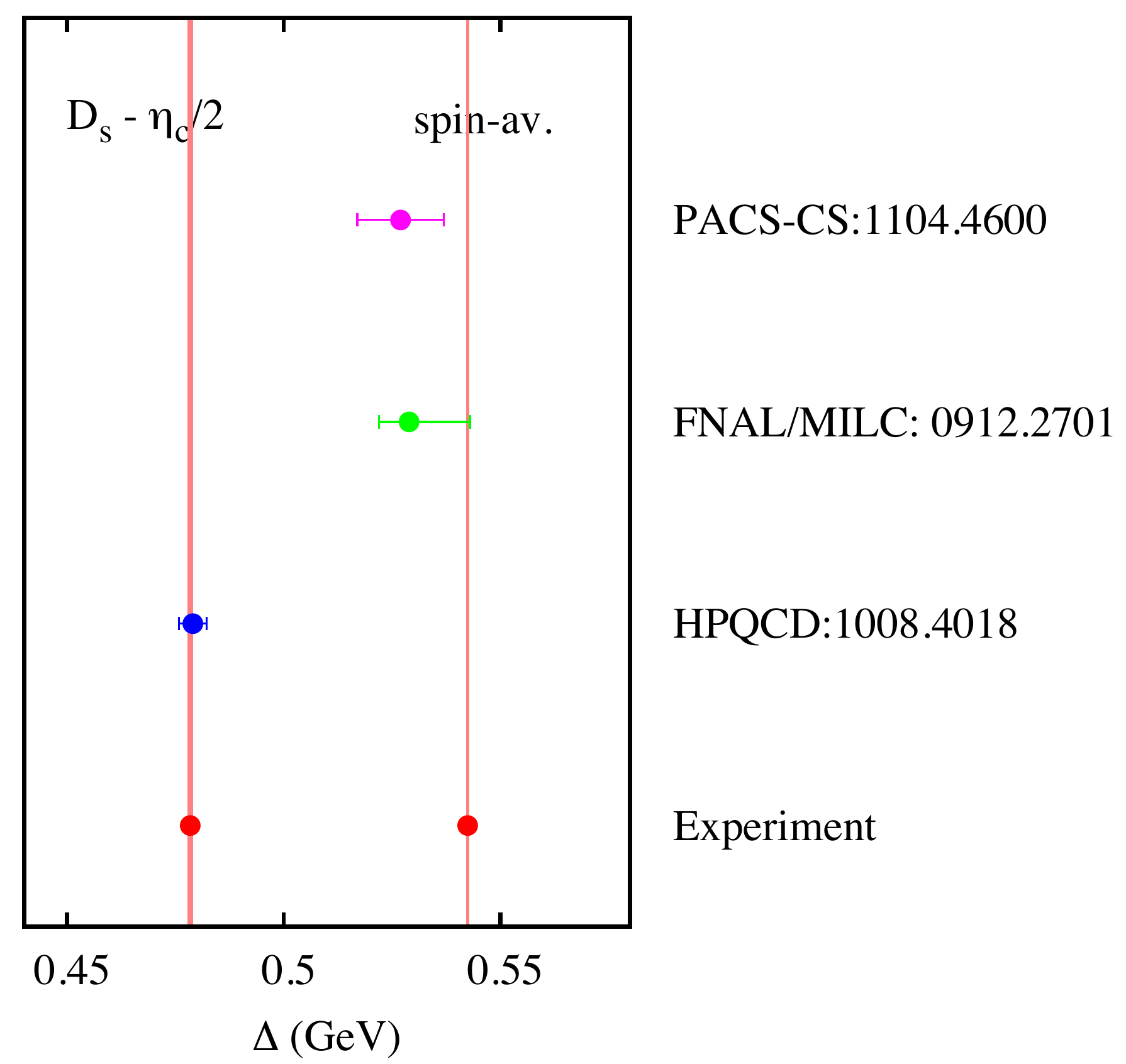}
}
\parbox{0.5\hsize}{
$b)$ \includegraphics[width=0.9\hsize]{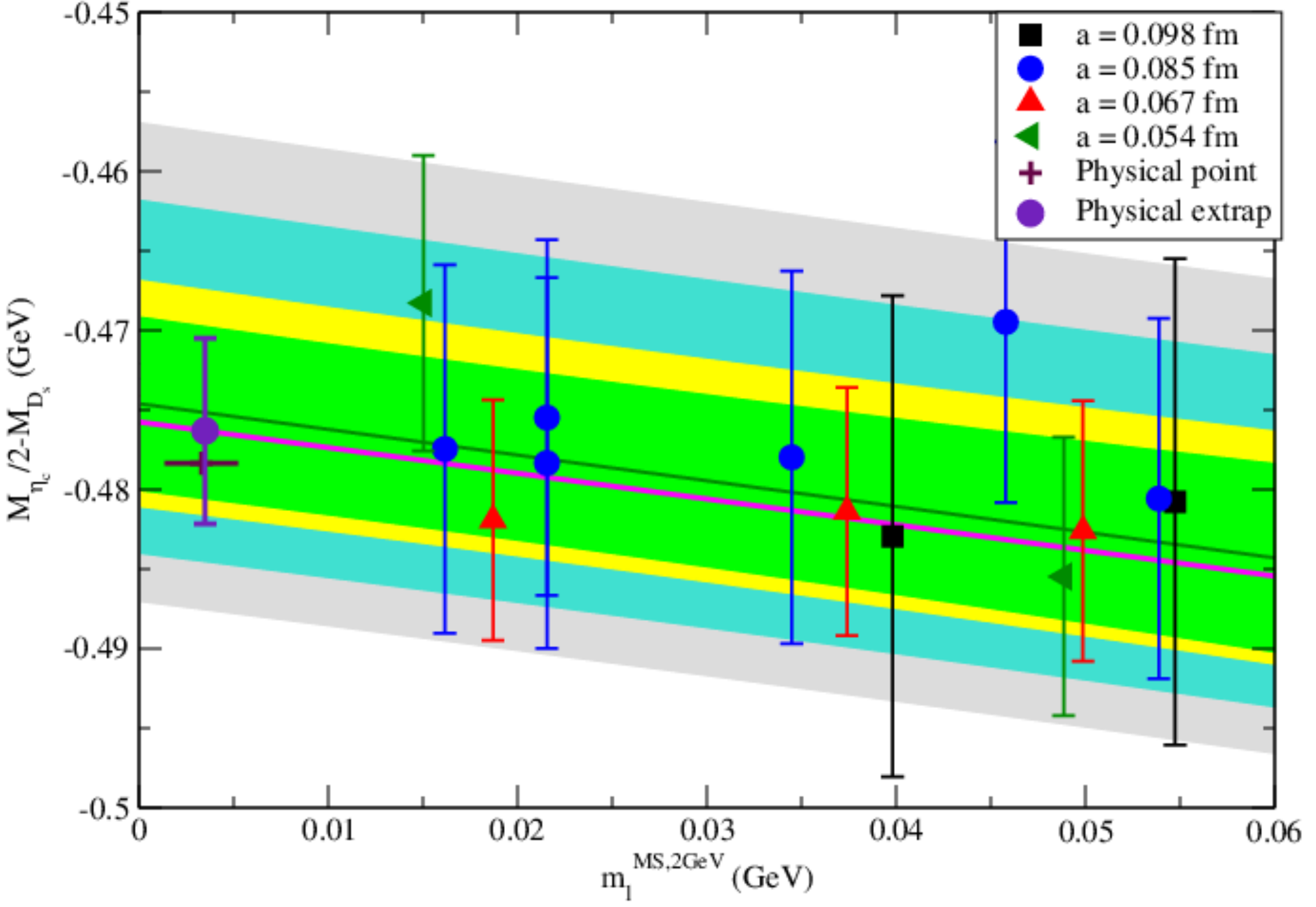}
}
\caption{ a) Comparison of lattice QCD 
results including $u$, $d$ and $s$ sea quarks
~\cite{pacscsspect, fermispect, fds} to experiment for 
the mass difference $\Delta = m(D_s)/m(\eta_c)/2$ or 
(on the right) a spin-averaged version of it.  
b) Results from ETMC for $\Delta$ on $n_f=2$ configurations
with a chiral and continuum extrapolation 
(marked as physical extrap.)~\cite{sanfill} compared 
to experiment (marked as physical point). 
}
\label{fig:mdscomp}
\end{figure}

The calculations above are fairly general ones aimed at 
mapping out a large part of the spectrum. I now want to 
turn to the `precision calculations' that are being done 
to provide weak decay rates for determination of the 
CKM matrix. These calculations focus on ground state mesons only 
and accuracy is critical. I will argue that accuracy must 
still be judged in tandem with results for the spectrum and 
other quantities that can be compared with experiment or between calculations. 

One variable that provides an excellent test in charm physics 
is the difference in mass, $\Delta$, between the $D_s$ meson and one half 
of the mass of the $\eta_c$. Typically one of these meson masses is used 
to fix the $c$ quark mass and the other is then a parameter-free 
determination from the lattice QCD calculation (assuming that the 
$s$ quark is determined from a light meson such as the $K$). 
HPQCD use the $\eta_c$, Fermilab/MILC use the $D_s$ and ETMC 
have tried both. 
$\Delta$ can be calculated in either case and is much more 
precisely determined than the masses themselves. This is 
because the difference is much 
smaller than either mass and so the 
absolute size of lattice spacing errors is 
much reduced. An error of a few percent is readily achieved in 
modern lattice QCD calculations and this would be impossible with 
any other method. $\Delta$ is the difference in binding energy 
between a heavy-light and a heavy-heavy meson and there are no 
good approximate treatments that apply to both systems. Certainly 
there are none that could be relied on for errors of a few percent.

Fig.~\ref{fig:mdscomp} summarises results 
for $\Delta$ from different calculations. 
The Fermilab Lattice/MILC result~\cite{fermispect} includes 
vector meson masses 
to make a `spin-averaged' mass difference. This 
removes a systematic error of $\mathcal{O}(\alpha_sa)$ 
resulting from the clover term in their action. Their total 
error is around 2\% (10 MeV),  dominated on the upward side by 
lattice spacing uncertainties from an old determination 
which has now been improved. 
The PACS-CS collaboration~\cite{pacscsspect} 
using a similar RHQ action also give a spin-averaged splitting. 
The PACS-CS result is 
from a calculation at one value of the lattice spacing only, with 
no estimate of lattice spacing errors. 
The HISQ result from HPQCD~\cite{fds} is much more accurate, with errors of 
less than 1\% (3 MeV). Small statistical errors enable the 
discretisation errors to be clearly identified and extrapolated away, 
underlining the advantages of a relativistic formalism. 
Note that the value and the error quoted by HPQCD includes an 
estimate of effects from electromagnetism, annihilation of the $\eta_c$, 
 and missing $c$ quarks 
in the sea. 

Fig.~\ref{fig:mdscomp} also includes results from the ETM collaboration using 
the twisted mass formalism for $c$ quarks on gluon 
configurations including only $u/d$ quarks in the sea. Since 
heavy-heavy mesons and heavy-light mesons are sensitive to 
different momentum scales it might be expected that $\Delta$ would
`see' the incorrect running of the strong coupling constant between 
scales that is a consequence of using only 2 flavors in the sea. 
Challenged to test this at the lattice conference ETM produced 
the right hand plot~\cite{sanfill} 
in fig.~\ref{fig:mdscomp} which shows that in fact a continuum 
and chiral extrapolation of $\Delta$ agrees well with experiment 
with errors of 2\% ((6)(4) MeV, where the first error is statistical and 
fitting and second an estimate of systematics). 

Using time moments of their statistically very precise $\eta_c$ correlators 
the HPQCD collaboration has developed a method with continuum QCD 
theorists~\cite{allison} that enables the $c$ quark mass to be extracted very 
accurately, using the high-order continuum perturbation theory 
that is available for the charm quark polarisation function. 
The result in the $\overline{MS}$ 
scheme, $\overline{m}_c^{(4)}(3\mathrm{GeV})$ = 0.986(6) GeV~\cite{curr}, 
agrees well with that determined 
from $e^+e^-$ cross-sections in the charm region using 
continuum methods. The ETM collaboration are now also applying this 
method and preliminary results presented here~\cite{etmmc} look promising. The 
$c$ quark mass is then another quantity that can be compared accurately 
between lattice QCD calculations (and with continuum results) and 
it would be good to have numbers from other formalisms. 
The accuracy in the $c$ mass can then be cascaded 
down to lighter masses using mass
ratios~\cite{mcs}. 

\subsection{Leptonic decays}
\label{sec:clept}
The annihilation rate of the charged $D$ and $D_s$ mesons to leptons 
via a $W$ boson is parameterised by the decay constant, $f_{D_{(s)}}$. 
This is defined (here for the $D_s$ at rest) as the matrix element 
between the meson and the vacuum 
of the temporal axial current that couples to the $W$ (the vector 
part of the $W$ interaction does not contribute): 
\begin{equation}
\langle 0 | \overline{c}\gamma_0\gamma_5 s | D_s \rangle = f_{D_s}M_{D_s}.
\label{eq:fds1}
\end{equation}
In a formalism with a partially conserved axial current, we can also use 
\begin{equation}
(m_s+m_c) \langle 0 | \overline{c}\gamma_5 s | D_s \rangle = f_{D_s}M_{D_s}^2.
\label{eq:fds2}
\end{equation}
The decay constant is a property of the meson related 
to the internal configuration of its valence quark and antiquark
 affected by their strong 
interaction. It is typically calculated in lattice QCD from the 
amplitudes in the same multiexponential fit to the meson 
correlator that gives the meson masses from the exponents. 
When the same operator, $\mathcal{O}$, is used to create and 
destroy the meson, the fit function for the correlator is: 
\begin{equation}
C_{2pt} = \sum_i a_i^2 f(E_i, t); \,\,\,\, f(E_i,t) = e^{-E_it} + e^{-E_i(T_p-t)}.
\label{eq:2pt}
\end{equation}
Here $E_i$ are the energies of different radial 
excitations - the ground state will be denoted $E_0$. 
$T_p$ is the time length of the lattice and the 
form of the time dependence allows for the meson to go 
round the lattice either way. For mesons containing staggered 
quarks there are typically additional oscillating terms that 
must be fitted. The amplitudes 
$a_i = \langle 0 | \mathcal{O} | i \rangle/\sqrt{2E_i}$, allowing the 
decay constants to be extracted from eqs.~\ref{eq:fds1} and~\ref{eq:fds2} 
if appropriate local operators are used for $\mathcal{O}$. 
The decay constants 
we will discuss here are all for ground-state mesons and therefore
extracted from $a_0$. 
For formalisms with a PCAC relation (such as HISQ or twisted 
mass) the decay constant is 
absolutely normalised. For formalisms without this level of 
chiral symmetry (such as the clover formalism) 
a renormalisation factor must be calculated 
to convert the lattice decay constant to a continuum value. 

The branching fraction for $D_s$ leptonic decay is then given by: 
\begin{equation}
\mathcal{B}(D_s \rightarrow l \nu_l) = \frac{G_F^2|V_{cs}|^2\tau_{D_s}}{8\pi} f_{D_s}^2m_{D_s}m_l^2\left( 1 - \frac{m_l^2}{m_{D_s}^2} \right)^2 .
\label{eq:leptrate}
\end{equation}
This has been determined by the BaBar, Belle and CLEO-c experiments 
and in both $\mu$ and $\tau$ modes.  Given a value for $V_{cs}$, for example from assuming unitarity of the CKM matrix, then an 
experimental value for $f_{D_s}$ can be extracted. 
Alternatively the comparison of theory and experiment can be used 
to determine $V_{cs}$. 

\begin{figure}[h]
\parbox{0.5\hsize}{
$a)$ \includegraphics[width=0.9\hsize]{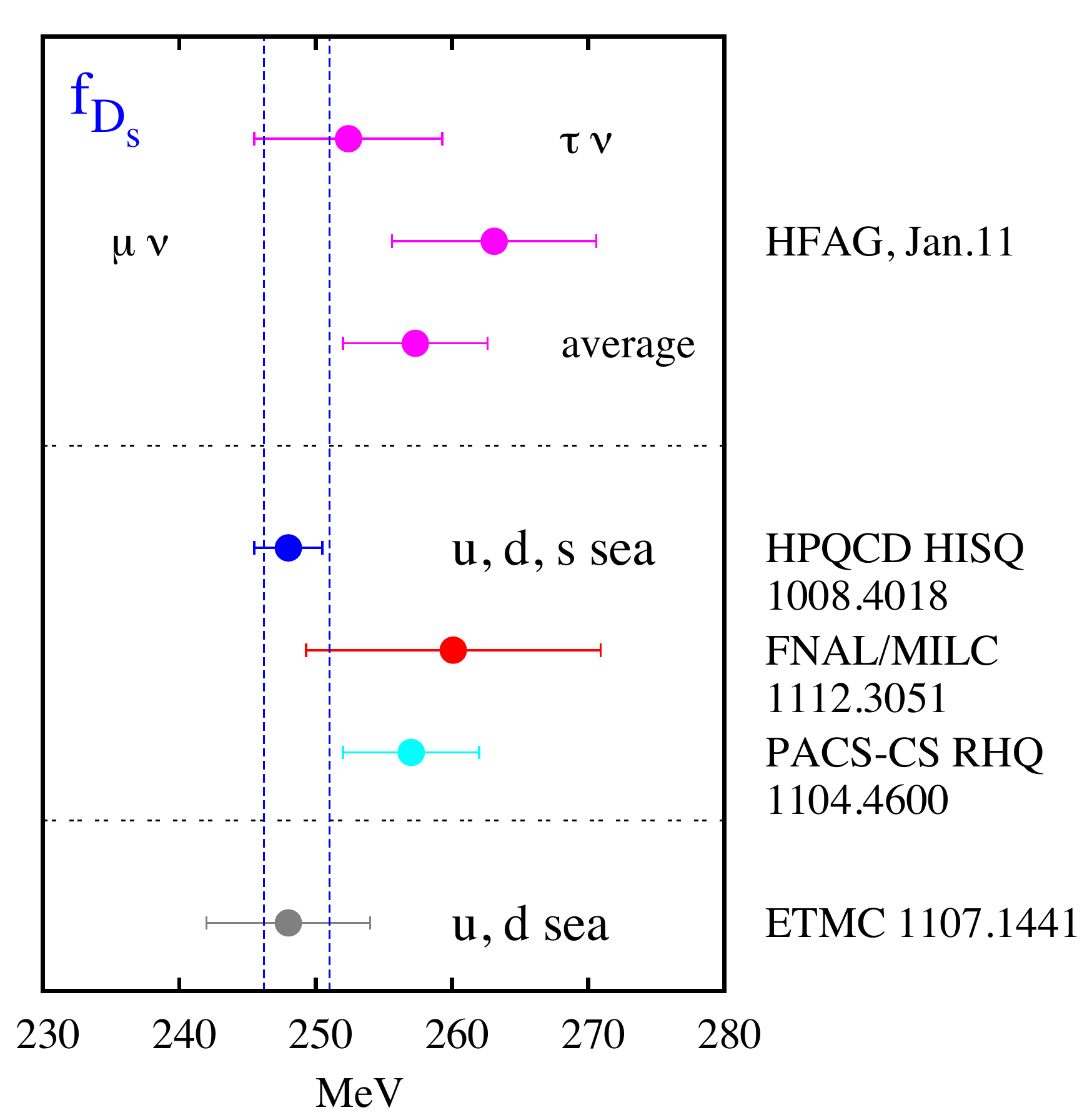}
}
\parbox{0.5\hsize}{
$b)$ \includegraphics[width=0.9\hsize]{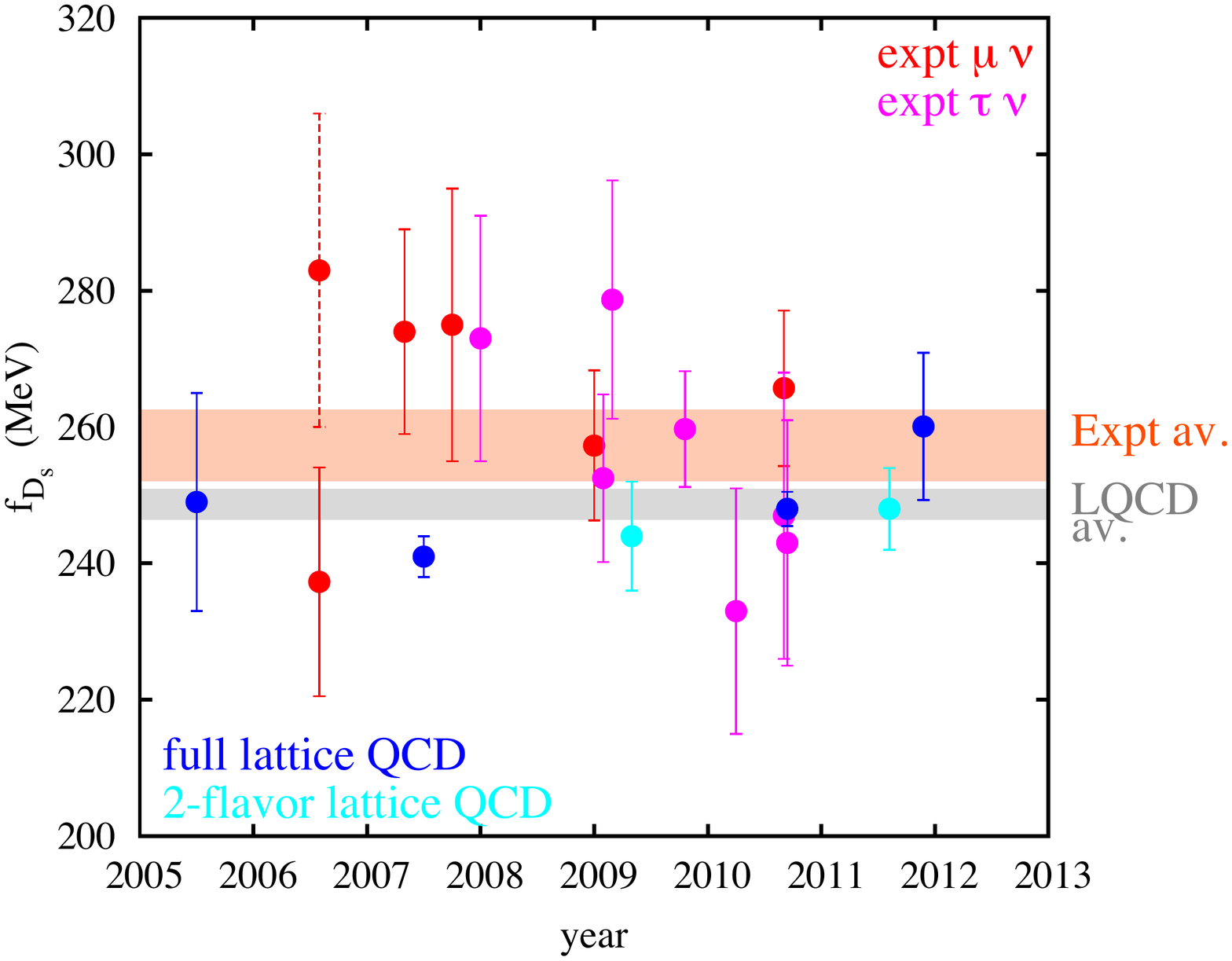}
}
\caption{ a) Comparison of lattice QCD results for $f_{D_s}$ to experimental 
averages extracted from leptonic decay rates (also showing $\mu \nu$ 
and $\tau \nu$ modes separately) using $V_{cs}=0.97345(16)$~\cite{hfag}. The lattice 
QCD world average is 248.6(2.4) MeV, shown with dashed lines; the experimental average is 257.3(5.3) MeV. 
b) A `history' plot for experimental and lattice results for $f_{D_s}$, updated from~\cite{fds}. 
}
\label{fig:fdscomp}
\end{figure}

Fig.~\ref{fig:fdscomp}a shows the current 
results for $f_{D_s}$ from both lattice 
QCD and experiment. The experimental averages are 
from January 2011~\cite{hfag}, but there have
been no new experimental results since. The experimental average over all modes gives $f_{D_s}$ = 257.3(5.3) MeV, using $V_{cs} = 0.97345(16)$. 

$f_{D_s}$ determination was specifically targeted 
by the CLEO-c and the B factory experiments 
as a good test of lattice QCD and this is why the 
subject has been vigorously pursued both by experimentalists 
and lattice QCD theorists. Because the $D_s$ 
has no valence light quarks, $f_{D_s}$ is relatively 
insensitive to the extrapolation of $m_{u/d}$ to the 
physical point and so the key issue for an accurate 
result is how well the valence charm quarks can be handled. 
Full lattice QCD results for $f_{D_s}$ 
date back to 2005 when a prediction of 249(16) MeV was 
given by the Fermilab Lattice/MILC collaboration ahead of 
experimental results starting in 2006. They used the Fermilab 
action for charm quarks and the asqtad improved staggered action 
for the strange quark. 
We have seen from the discussion in the introduction that 
nonrelativistic methods for heavy quarks reduce
discretisation errors. They have several disadvantages, however, 
and I would claim that it was fairly conclusively demonstrated by 
the HPQCD collaboration in 2007~\cite{oldfds} that an improved relativistic 
action (HISQ) was superior. HPQCD's result for $f_{D_s}$ had a 1.5\% 
error, similar to those possible in the same calculations for the light decay 
constants $f_K$ and $f_{\pi}$. This is not surprising - statistical 
and scale uncertainty errors are fairly similar 
in all cases, there is no normalisation uncertainty, 
and the chiral extrapolation 
for $f_K$ and $f_{\pi}$ is over a similar range to the $a$ extrapolation 
of the $D_s$ results. 

In 2008 the experimental results for $f_{D_s}$ were around 275 MeV, 
although individual results had fairly large errors of 15-20 MeV. 
This gave a very exciting picture for a while, as mapped out in the 
history plot in fig.~\ref{fig:fdscomp}b. 
Although the 3$\sigma$ discrepancy between lattice QCD and experiment 
that existed then 
has now fallen to 1.6$\sigma$~\cite{fds} as the experimental values moved down 
and the lattice QCD result moved up, it showed very clearly how 
precision lattice QCD calculations can have an impact on experiment.  

The current picture is summarised in fig.~\ref{fig:fdscomp}. 
The HPQCD result was updated to 248.0(2.5) MeV in 2010~\cite{fds} with a 
recalibration of the lattice spacing. This now includes results 
from MILC asqtad $n_f=2+1$ configurations at 
5 values of the lattice spacing from 0.15fm down to 0.04fm and 
multiple $m_{u/d}$ values. 
Absolute normalisation of the decay constant 
 is also possible in the twisted mass formalism and ETMC gave 
a result of 244(8) MeV in 2009~\cite{etmoldf} using $n_f=2$ twisted mass configurations. 
This was updated in 2011 to 248(6) MeV~\cite{tmfb} using 4 values of the lattice 
spacing from 0.1fm down to 0.05fm and multiple $m_{u/d}$ values. 
In 2011 The PACS-CS collaboration gave 257(5) MeV~\cite{pacscsspect}, 
including only a partial estimate of errors.  
They used an RHQ action at one value of the lattice 
spacing with $n_f=2+1$ flavors of clover sea quarks, 
but having reweighted a small ensemble of configurations 
to the physical $m_{u/d}$ point. 

The Fermilab Lattice/MILC collaboration have also recently updated their 
analysis~\cite{fnalnewf} using the 
Fermilab action on the 0.09 and 0.12 fm MILC 
2+1 asqtad lattices. They obtain an improved result of 260.1(10.8) 
MeV where the error is dominated by heavy quark effects, i.e. 
estimates of the mixed relativistic/discretisation corrections 
from matching the Fermilab action to continuum QCD. 
The error from the overall renormalisation constant, $Z_{A_{Qq}^4}$, 
to be applied 
to convert from the lattice decay constant to the continuum one 
is taken to be 1.5\%. The renormalisation is done using 
a method which combines perturbative 
and nonperturbative techniques. $Z$ for
the local temporal vector current can be determined nonperturbatively 
in both the heavy-heavy case 
and the light-light case using the 
normalisation condition: 
\begin{equation}
1 = \langle H_{q\overline{q}} | Z_{V^4_{qq}} | H_{q\overline{q}} \rangle .
\label{eq:vecnorm}
\end{equation}
Then the temporal axial current renormalisation needed for $f_{D_s}$ is defined by: 
\begin{equation}
Z_{A^4_{Qq}} = \rho_{A^4_{Qq}} \sqrt{Z_{V^4_{qq}}Z_{V^4_{QQ}}}
\label{eq:zrat}
\end{equation}
and $\rho_{A^4_{Qq}}$ is calculated through $\mathcal{O}(\alpha_s)$ in 
lattice QCD perturbation theory. A surprising and significant result 
is found - the coefficient of $\alpha_s$ is very small (< 0.1, but 
not zero) for 
heavy clover quark masses up to around $ma=1$  after which it 
rises linearly with $ma$. 
This is shown in Fig. 3 of~\cite{fnalz} but note that this is a plot of 
the coefficient of $g^2$. The $x$-axis is the quark mass 
$m_1a=\log(1+m_0a)$ where $m_0a$ is the standard mass. 
The relevant region, even for $b$ quarks, is then quite restricted:
$m_0a=3$ corresponds to $m_1a =1.4$. 
It is not clear why the $\alpha_s$ coefficient in $\rho_{Qq}$ is so small.  
Some cancellation of perturbative corrections 
between $Z_V$ and $Z_A$ in eq.~\ref{eq:zrat} 
seems reasonable but it would be good to understand if the tiny 
one-loop coefficient near clover masses of zero is accidental or 
whether it is telling us something. 
Does it hold for the combination of clover quarks with other 
light quarks in general? It seems to hold, but to a somewhat lesser 
extent, for other types of 
current, such as the spatial vector~\cite{fnalz}. 
It is not clear to me what this says about the next order in
perturbation theory and it would be good to see some nonperturbative 
tests of this. A simple test would be to compare the amplitudes
of mixed clover-staggered pseudoscalar correlators to those of 
absolutely normalised 
staggered-staggered correlators at the same mass, for example 
that of the $\eta_s$. At these lower masses discretisation errors 
are less of an issue and a connection could be made to calculations 
using relativistic clover quarks for light decay constants, which 
could be interesting ( for a review of issues here see~\cite{wittig} ).   
Fermilab/MILC take an error on $f_{D_s}$ from uncalculated higher orders
in perturbation theory in $\rho_{A^4_{Qq}}$ 
of 0.1$\alpha_s^2$. This seems optimistic 
to me without further tests, but on the other hand this is academic at 
present because even a much more pessimistic error would 
have little impact on the total error. 

The current HPQCD and Fermilab/MILC results supersede their 
earlier results and are the only two 
using 2+1 flavors of sea quarks with a complete error 
budget. Even though they both use the MILC configurations, there 
is little overlap in the key ensembles in the two cases, so 
I take their errors to be independent. 
I then combine them into a world-average value of $f_{D_s}$ 
from lattice QCD of 248.6(2.4) MeV, not surprisingly dominated by 
the HPQCD result. 

In the ratio $f_{D_s}/f_D$ 
$Z$ factors cancel and so the errors are similar at 2-3\% between 
HPQCD results using HISQ (1.164(18)~\cite{fds}, updating the 
error from~\cite{oldfds}) 
and Fermilab/MILC (1.188(25)~\cite{fnalnewf}). This gives 
a world average of 1.172(15) where I have now taken a 100\% correlation
between the statistical errors of the two calculations since the 
$f_D$ calculations used the same ensembles. 
This value is not yet accurate enough to tell whether it 
agrees or disagrees with the similar ratio $f_K/f_{\pi}$ which has 
been calculated to be 1.193(5) 
from lattice QCD~\cite{wittig}. Note also that $f_{\eta_s}/f_K$, 
another ratio in which the numerator and denominator differ by 
the substitution of an $s$ quark for a $u/d$, is 
1.165(8) from lattice QCD~\cite{r1, newups}.  Surprisingly the 
PACS-CS result of $f_{D_s}/f_D$ = 1.14(3)~\cite{pacscsspect}, which 
used $m_{u/d}$ values at the physical point, is lower in its central 
value than either HPQCD or Fermilab/MILC, who have to extrapolate 
upwards to that point. I have not included this 
number in the average since it was only obtained on configurations 
at one lattice spacing. 
To improve results for $f_{D_s}/f_D$ clearly needs more chiral 
lattices; there is no particular need for finer lattices. The new 
sets of lattices with $u/d$ sea quark masses at the chiral 
point which are being generated 
should allow $f_{D_s}/f_D$ to be calculated with 1\% accuracy. 
The experimental average for $f_{D_s}/f_D$ is poorly determined 
at 1.26(5)~\cite{rosnerstone} because there is no cancellation of errors 
in the ratio. The experimental result for $f_D$ is 206.1(8.9) MeV from 
CLEO-c~\cite{rosnerstone} using
$V_{cd}$ from CKM unitarity = 0.2252(7)~\cite{pdg}. 
Combining the average lattice $f_{D_s}$ with the average 
ratio of $f_{D_s}/f_D$ gives $f_D$ = 212.1(3.4) MeV. 

\subsection{Semileptonic decays}
\label{sec:csemil}

\begin{figure}[h]
\parbox{0.5\hsize}{
$a)$ \includegraphics[width=0.9\hsize]{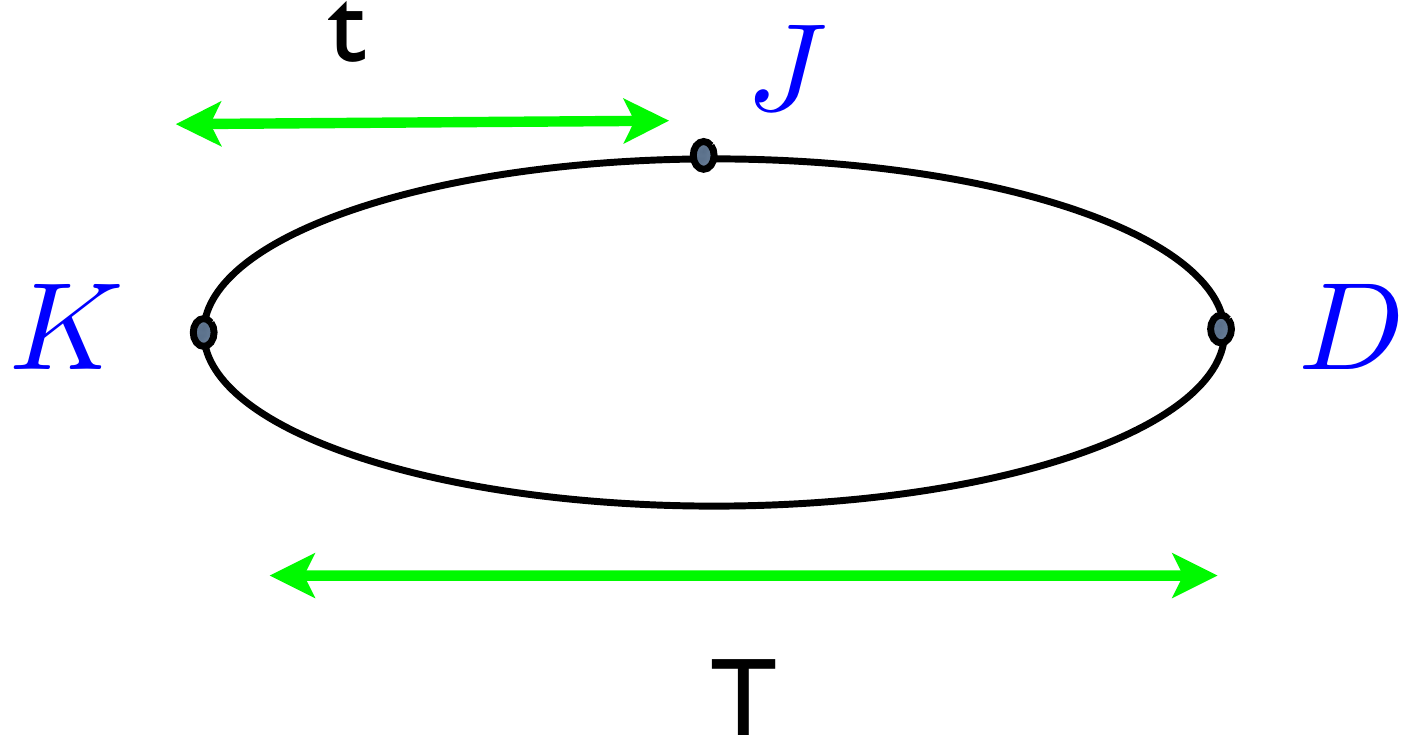}
}
\parbox{0.5\hsize}{
$b)$ \includegraphics[width=0.9\hsize]{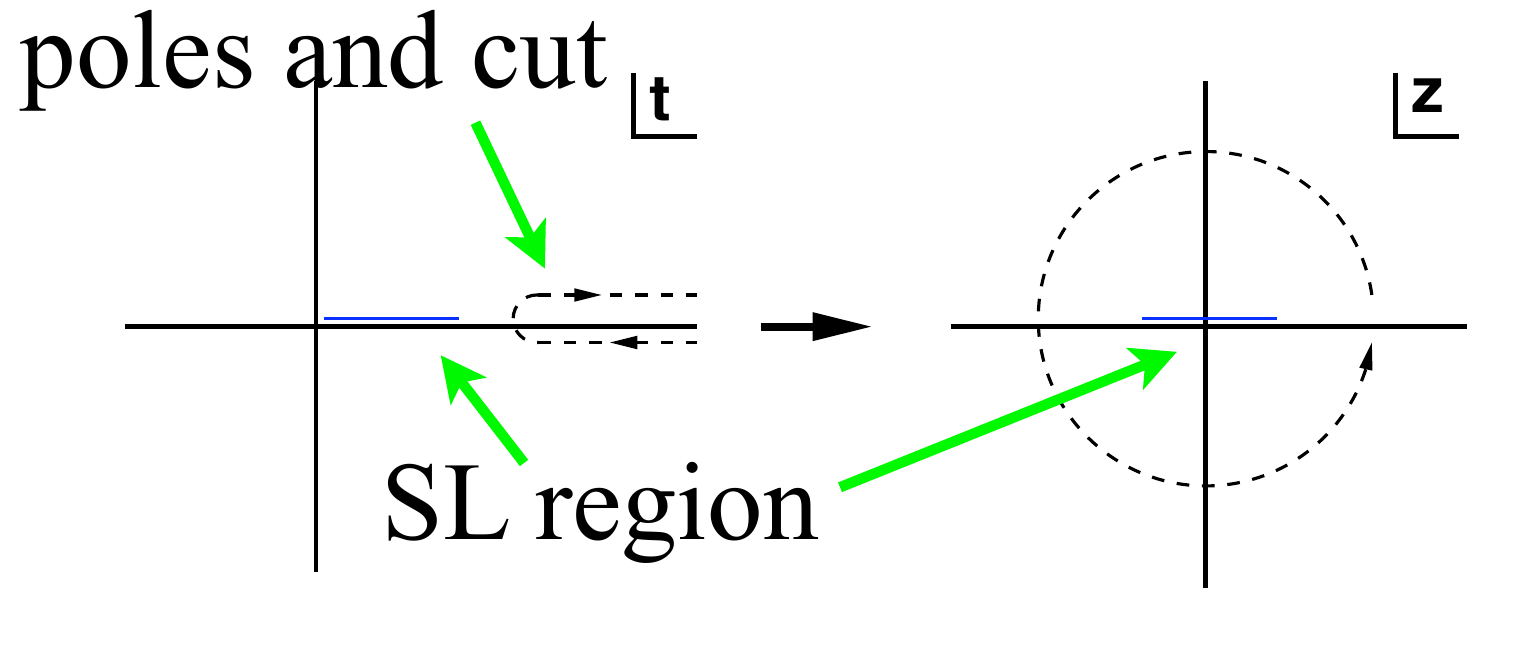}
}
\caption{ a) Sketch of a 3-point function. The operators 
for $D$ and $K$ are a distance $T$ apart on the lattice and 
the current operator is inserted between them. The 3-point 
correlator should be calculated for all $t$ from 0 to $T$ and for several 
different $T$ values.  
b) Transformation of the complex $t=q^2$ plane to the $z$ plane for 
semileptonic form factors. 
}
\label{fig:3pt}
\end{figure}

The analysis of semileptonic decays in which one meson changes into another 
and emits a $W$ boson gives us access to more detailed 
information about meson internal structure than the one number 
represented by the decay constant. The information from 
QCD in these decay processes is parameterised by form 
factors that are functions of 
$q^2$, the square of the 4-momentum transfer from the initial 
to the final meson. Calculation of the form factors in 
lattice QCD allows the $q^2$-dependence of the rate for such 
decays to be compared to experiment as well as CKM 
matrix elements to be extracted. Since the lattice calculation 
corresponds to a specific final state meson, the experimentalists 
must identify this meson in their sample of decays to 
give the `exclusive' (as opposed to the `inclusive') rate.  
The simplest case is that of pseudoscalar 
to pseudoscalar decay in which only the vector piece of 
the current coupling to the $W$ contributes, and there are 
two form factors, the vector form factor $f_+$ and the 
scalar form factor $f_0$. Illustrating 
this for $D \rightarrow K l \overline{\nu}$ decay we have: 
\begin{equation}
\langle K | V^{\mu} | D \rangle = f_+(q^2) \left[ p_D^{\mu} + p_K^{\mu} - \frac{m_D^2 - m_K^2}{q^2} q^{\mu} \right] + f_0(q^2) \frac{m_D^2-m_K^2}{q^2}q^{\mu} .
\label{eq:ff}
\end{equation}
The matrix element on the left-hand side is determined from the 
amplitude of a "3-point correlator" sketched in Fig.~\ref{fig:3pt}. The source 
and sink mesons are taken a distance $T$ apart on the lattice and 
the vector current is inserted a distance $t$ from the source. $t$ is 
allowed to run over all values from $0$ to $T$ and the correlator 
is fit simultaneously as a function 
of $t$ and $T$ along with the standard "2-point" 
meson correlators for the source and sink mesons, for which 
the fit function is given in eq.~\ref{eq:2pt}. 
The 3-point function fit is :
\begin{equation}
C_{3pt} = \sum_{i,j} a_i b_j V_{ij} f(E_{a,i},t) f(E_{b,j},T-t)
\label{eq:3pt}
\end{equation}
where $a_i$ and $b_j$ are the same amplitudes and $E_{a,i}$ and 
$E_{b,j}$ are the same energies as in the 2-point function fits 
for mesons $a$ and $b$ respectively
and $V_{ij}$ is the matrix element of the vector current between 
them. The result we typically want, and that I will discuss here, 
 is that between the ground states, $V_{00}$. When staggered quarks 
are used there are generally additional 
oscillating pieces that need to be fit.  

To cover the range of $q^2$ values in the decay one or both of the 
source or sink mesons can be given a spatial momentum. It is simplest 
to work in the rest frame of the source meson (for example the $D$ above). 
Then, when the $K$ is at rest, $q^2$ is a maximum at $(m_D-m_K)^2$, 
and the lepton and antineutrino emerge back-to-back. This is the 
easiest kinematics to reproduce on the lattice, but the experimental 
rate at this point is zero (see eq.~\ref{eq:3ptrate} below).
The other extreme is $q^2=0$ when the lepton and antineutrino balance 
the $K$ momentum. Lattice QCD errors grow as spatial momentum of the 
$K$ increases. The experimental errors are best for small $q^2$ 
(but not necessarily 0)
where the rate is larger. 

When the $W$ decay to leptons is folded in with eq.~\ref{eq:ff}, 
the contribution to the rate 
from $f_0(q^2)$ appears multiplied by lepton masses and so for $e$ and $\mu$ 
channels is negligible.  Then :
\begin{equation}
\frac{d\Gamma}{dq^2} = \frac{G_F^2p_K^3}{24\pi^3} |V_{cs}|^2 |f_+(q^2)|^2 .
\label{eq:3ptrate}
\end{equation}
There is a useful kinematic constraint from eq.~\ref{eq:ff} at $q^2=0$: $f_+(0)=f_0(0)$. This is helpful because the scalar form factor can be absolutely 
normalised from the PCVC relation $\partial_{\mu}V^{\mu} = (m_1-m_2)S$ when 
the same formalism is used for the two quarks in the weak decay ($c$ and 
$s$ for $D \rightarrow K$ decay) and 
they differ in mass. Then: 
\begin{equation}
\langle K | S | D \rangle = f_0(q^2) \frac{m_D^2-m_K^2}{m_c-m_s} ,
\label{eq:ffscalar}
\end{equation}
where $m_c$ and $m_s$ are the bare lattice quark masses. This method 
has recently been introduced by the HPQCD collaboration~\cite{na1}, reducing 
lattice errors significantly for $f_+(0)$. 

Continuum theorists have developed a good understanding of the behaviour 
of form factors based on their pole and cut structure in the wider 
complex $q^2$ plane (see, for example,~\cite{lellouch}). 
The $q^2$ region for semileptonic decay runs 
from $q^2=0$ up to (for $D \rightarrow K$) $q^2_{max} = t_- = (m_D-m_K)^2$.  
Above $t_+=(m_D+m_K)^2$ where a real $D$ and $K$ can be exchanged, 
there is a cut. In addition there may be poles if there are isolated 
resonances with the right quantum numbers with masses between $t_-$ 
and $t_+$. For example, the $D\rightarrow K$ vector form factor 
has a pole at $q^2=m_{D_s^*}^2$ where $m_{D_s^*} - m_{D}$ = 243 MeV 
and the scalar form factor has a pole at $m_{D_{s0}} - m_D$ = 448 MeV 
(ignoring the suppressed $D_s\pi$ cut). 
The form factor diverges at the pole, but this is outside the 
physical region for semileptonic decays so what is seen is a rise 
of the form factor as $q^2$ increases to $q^2_{max}$. 

It is useful to remove this pole behaviour from the form factor 
so that $\tilde{f}(q^2 ) = f(q^2)/(1-q^2/m_{pole}^2)$ and 
then to transform $\tilde{f}$ into $z$-space where
\begin{equation}
z = \frac{\sqrt{t_+-q^2}-\sqrt{t_+-t_0}}{\sqrt{t_+-q^2}+\sqrt{t_+-t_0}} .
\label{eq:z}
\end{equation}
Eq.~\ref{eq:z} maps the line around the cut 
from $q^2 = \infty$ to $q^2=t_+$ and back 
again to the circle with $|z|=1$, as shown in Fig.~\ref{fig:3pt}b. 
The physical 
semileptonic region is then inside this circle, where $\tilde{f}$
is finite and well-behaved. The exact position 
of the physical region depends on the value 
for $t_0$ (since $z(q^2=t_0) =0$). Often the region is made 
symmetric about $z=0$ to minimise the maximum $z$ value but it 
can be more convenient to take $t_0=0$. The form 
factor $\tilde{f}$ in the physical region can be described 
by a simple power series in $z$. Allowing 
the coefficients of the expansion to depend on lattice spacing and 
$m_{u/d}$ provides a straightforward way to extrapolate to the 
continuum and chiral limits which, for example, 
reduces the confusion between $q^2$ dependence 
and $m_{u/d}$ dependence. The continuum and chiral limit 
form factor $f(q^2)$ can then easily be reconstituted at the end.  

Lattice QCD calculations for semileptonic form factors have improved 
hugely over the last few years as techniques have developed. State-of-the-art 
calculations now have: 
high statistics, including the use of random 
wall sources to improve this further; 
multi-exponential fits to the 3-point function for multiple $T$ 
values and not simply a "plateau" search in a ratio of 3-point and 
2-point functions; use of a phase rotation of the gluon field to 
tune spatial momenta accurately to find the $q^2=0$ point~\cite{twist} and use 
of the $z$ expansion~\cite{fnalvis, na1} as described above to fit the 
form factor shape and improve 
extrapolation to the chiral/continuum limit.  

\begin{figure}[h]
\parbox{0.45\hsize}{
$a)$ \includegraphics[width=0.9\hsize]{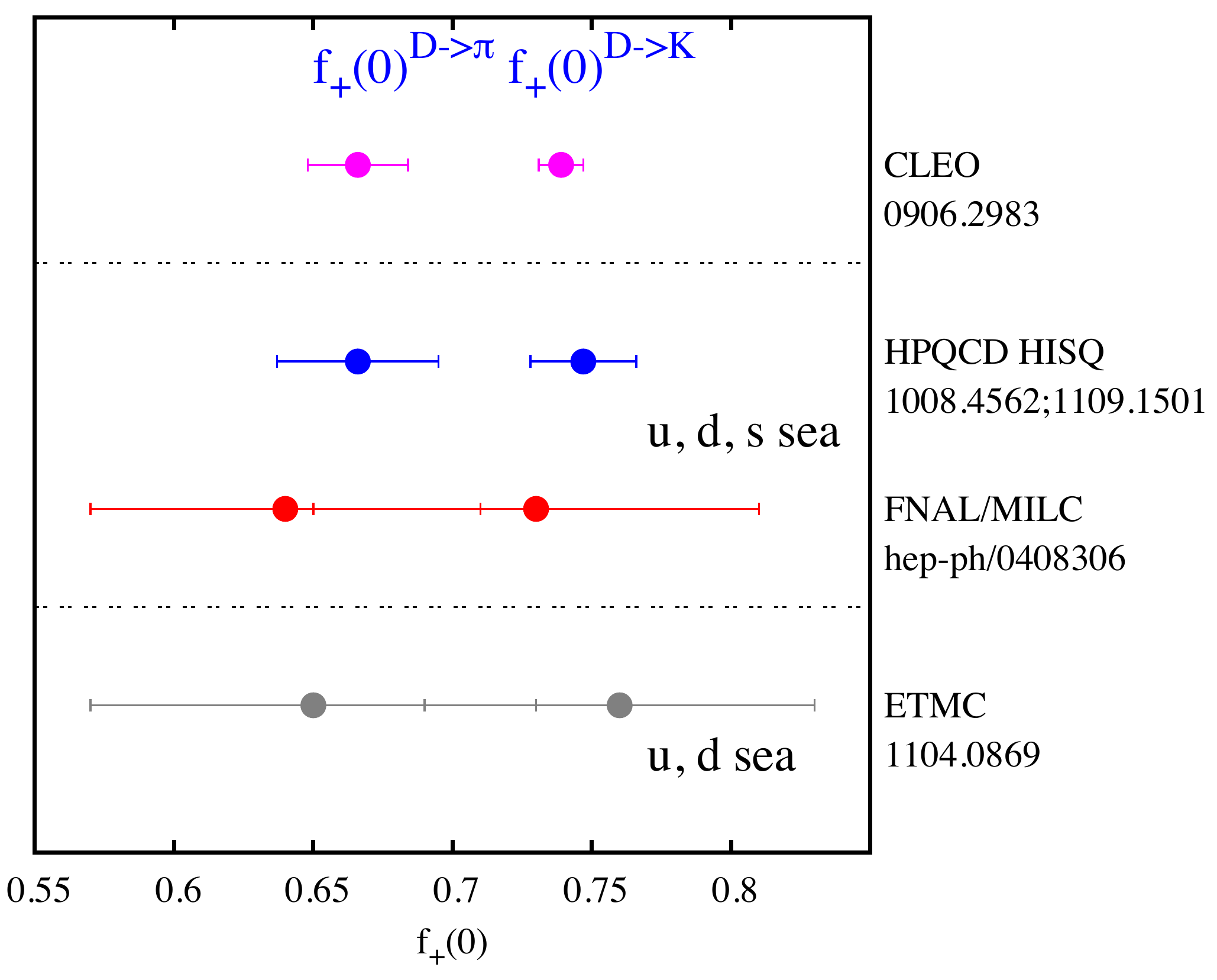}
}
\parbox{0.55\hsize}{
$b)$ \includegraphics[width=0.44\hsize]{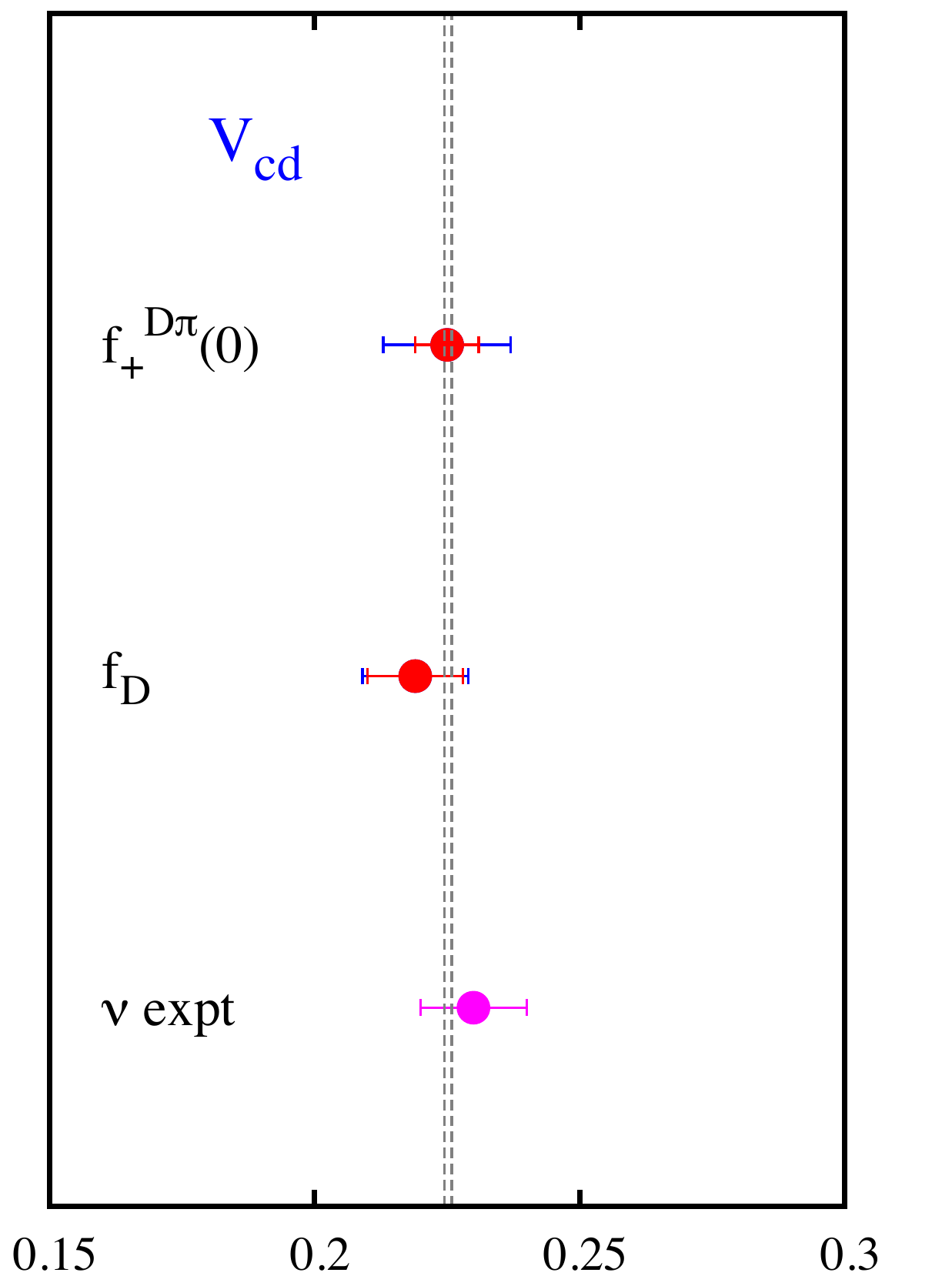}
\includegraphics[width=0.4\hsize]{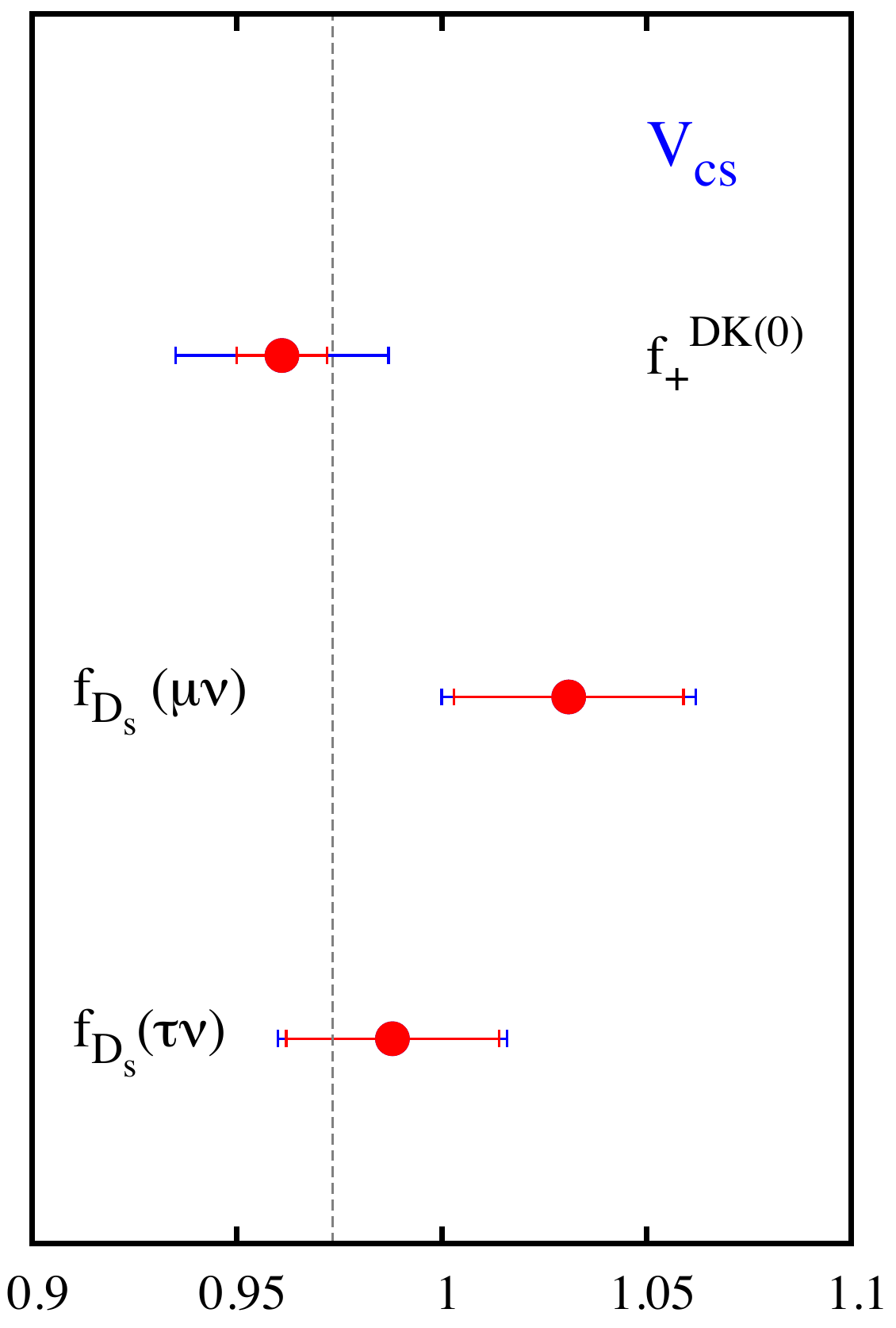}
}
\caption{ a) Lattice QCD results for $f_+(0)$ from $D \rightarrow K$ 
and $D \rightarrow \pi$ decay, compared to results from 
fitting experimental semileptonic rates as a function of $q^2$~\cite{cleoff}, 
determining $f_+(0)V_{cx}$ and then taking 
$V_{cs}$=0.97345(16) and $V_{cd}$=0.2252(7). 
b) $V_{cd}$ and $V_{cs}$ determined from combining results for 
semileptonic and leptonic decay rates with lattice QCD world 
average values for form 
factors and decay constants given in the text. 
The inner (red) error bar is from experiment; 
the outer error bar is the total error bar including that 
from lattice QCD. The $V_{cd}$ plot includes the direct determination 
from neutrino experiments~\cite{pdg}. The dashed grey lines 
give the values from assuming CKM unitarity~\cite{pdg}.   
}
\label{fig:3ptres}
\end{figure}

Experimentalists often quote their results, for example 
for $D \rightarrow K$, in the form of a value 
for $V_{cs}f_+(0)$, obtained from a fit to their 
data which includes a parameterisation of the form factor.     
A lattice calculation of $f_+(0)$ can then be used to 
determine the CKM element. Fig.~\ref{fig:3ptres} shows the current 
status of lattice QCD calculations for $f_+(0)$ for 
$D \rightarrow K$ and 
$D \rightarrow \pi$, along with experimental results obtained 
by using unitarity values for $V_{cs}$ and $V_{cd}$. 
The HPQCD/HISQ results~\cite{na1, na2} are obtained from $f_0(0)$ which, 
as discussed above, is absolutely normalised.  
This shows a big improvement over earlier
results in that
$D \rightarrow K$ and $D\rightarrow \pi$ values
are clearly distinguished, with the accuracy improved to the 
level of a few percent. Since these are the only results in 
full QCD to date using more than one lattice spacing  we simply 
quote these numbers as the `world average': 
$f^{D \rightarrow K}_+(0)$ = 0.747(19) 
and $f^{D \rightarrow \pi}_+(0)$ = 0.666(29).  

Fermilab/MILC gave a progress report on their calculation 
of $D \rightarrow \pi$ $f_+(q^2)$ this 
year~\cite{bailey} showing results from 11 ensembles and a shape that agrees 
well with CLEO-c data. They project future errors 
on $f_+(0)$ of less than 5\%, dominated 
by heavy quark and chiral extrapolation errors. The renormalisation 
of the clover-staggered vector current is done in the same way 
as for decay constants discussed in section~\ref{sec:clept}.

HPQCD also presented new results on $f_+(q^2)$ obtained using 
a spatial vector current normalised in the symmetric case using 
eq.~\ref{eq:vecnorm} since the $Z$ factor does not depend significantly 
on mass~\cite{jonna}. This is found to agree with a local 
temporal charm-strange vector current 
which can be normalised by the fact that at $q^2_{max}$ the vector matrix 
element in eq.~\ref{eq:ff} is equal to $f_0(q^2_{max})(m_D+m_K)$.
The phase technique~\cite{twist} is used to 
tune across the physical $q^2$ range, 
including at $q^2=0$. $f_+$ and $f_0$ are fit together in $z$-space, 
using $t_0=0$ so that the constraint $f_+(0)=f_0(0)$ can be maintained.  
The final form factor shapes agree well with experiment and we 
expect a further factor of two improvement in the determination of 
$V_{cs}$ from this method. 

These results~\cite{jonna} showed up an interesting fact, not apparently 
noticed before, that the form factors for charmed meson 
decay are very insensitive 
to the spectator quark mass as it is varied between 
light and strange. This is within statistical errors of 1-2\% at 
high $q^2$ and 2-5\% at $q^2=0$. 
The insensitivity has significant consequences. One 
is that $D_s \rightarrow K$ and $D \rightarrow \pi$ form 
factors are the same~\cite{jonna}, which can be tested experimentally.  
Another is that this would then be expected to hold also for $B$ meson 
decays so that $B_s \rightarrow D_s$ and $B \rightarrow D$ 
form factors would be equal. This is useful for experimental 
normalisation and will be discussed further in section~\ref{sec:bother}. 

Further results this year came from QCDSF~\cite{qcdsf} looking at the 
disconnected contributions necessary 
to determine $D_{(s)} \rightarrow \eta^{(\prime)} l \nu$ rates 
(using a relativistic smeared clover action for all quarks)
and from HPQCD on axial and vector form factors for 
$D_s \rightarrow \phi l \nu$~\cite{gordon} (using HISQ).  
These form factors add to the range of quantities that can 
be determined from essentially the same lattice QCD calculation 
and tested against experiment with no free parameters. 

\subsection{$V_{cs}$ and $V_{cd}$}
\label{sec:vcsd}

Fig.~\ref{fig:3ptres}b shows the current status of the direct determination 
of $V_{cs}$ and $V_{cd}$ using leptonic decays of $D$ and $D_s$ 
mesons and semileptonic $D \rightarrow \pi$ and $D \rightarrow K$ 
decays, combining experimental results with lattice QCD world 
averages given in sections~\ref{sec:clept} and~\ref{sec:csemil}. 
The results will be summarised in terms of unitarity tests 
of the CKM matrix in section~\ref{sec:conclusions}. 

\section{Bottom physics results}

\subsection{Spectroscopy}
This year has seen progress on improving methods for $b$ physics. 
The HPQCD collaboration has now taken the NRQCD action to 
the next stage by determining the $\mathcal{O}(\alpha_s)$ 
corrections to the sub-leading ($v^4$) terms~\cite{hammant}. Improved analysis 
of the bottomonium spectrum was 
shown~\cite{newups, rachel} including the hyperfine 
splitting (the mass difference between the $\Upsilon$ and the 
$\eta_b$) and predictions for $D$-wave states~\cite{daldrop}. 
These calculations 
have been done on MILC configurations 
including $u$, $d$, $s$ and $c$ HISQ quarks in the sea 
and with a further 
improved gluon action. 
Heavy-light physics is underway on these 
configurations. The improvements to the action should reduce 
NRQCD systematic errors to the level of 5 MeV when determining 
the difference between, say, the $B_s$ meson mass and one 
half the mass of the $\Upsilon$.  
This is not enough to distinguish the effects of $c$ quarks in 
the sea since they are expected from a perturbative analysis~\cite{gregory} 
to shift the $\Upsilon$ 
by around 5 MeV but have 
little effect on the $B_s$. It might, however, be enough to see the 
effect of quenching $s$ quarks since, based on old results 
from quenching all 3 light quarks, this could affect this mass splitting 
at the 10-20 MeV level~\cite{gray}, as well as affecting radial 
and orbital $\Upsilon$ splittings 
at the 2-3\% level (relative to the 
$2S-1S$ splitting). 
It would be interesting to do a systematic analysis on $n_f=2$ 
configurations to see if an error in the comparison 
to experiment shows up here, since no 
effect has so far been seen anywhere else. This would help to 
quantify the consequences of quenching $s$ quarks, 
since these cannot be easily estimated.  

Both ETMC~\cite{tmfb} and HPQCD~\cite{fbs} have 
given new results, to be discussed 
further below, using relativistic 
approaches (twisted mass and HISQ respectively) 
for heavy quarks and extrapolating in the heavy 
quark mass up to the $b$. 
The RBC/UKQCD collaboration have tuned 
their RHQ action for $b$ quarks and are now using that action for 
bottomonium and heavy-light physics~\cite{witzel}. 

A useful comparison between lattice QCD calculations 
is in the determination of 
the $b$ quark mass. The Alpha~\cite{alphalat11} and 
ETMC~\cite{tmfb} collaborations presented 
new results on $m_b$ from configurations that include $u$ and $d$ 
quarks in the sea. 
Alpha use HQET through $1/m_b$ on the lattice and find $m_b^{(n_f=2)}(m_b)$ in 
the $\overline{MS}$ scheme to be 4.23(15) GeV 
in agreement with the ETMC result of 4.29(14) GeV. The ETMC result uses 
the twisted mass formalism for the heavy quarks, extrapolating up to the 
$b$ using a function with a known static limit. 

\begin{figure}[h]
\parbox{0.5\hsize}{
$a)$ \includegraphics[width=0.95\hsize]{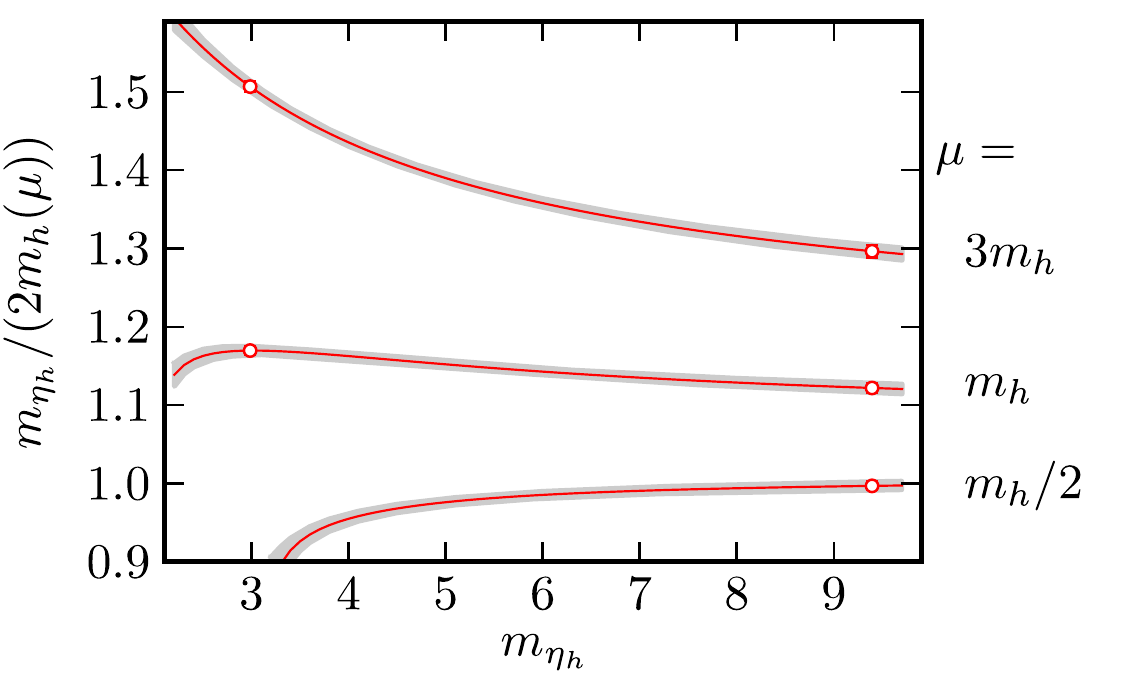}
}
\parbox{0.5\hsize}{
$b)$ \includegraphics[width=0.95\hsize]{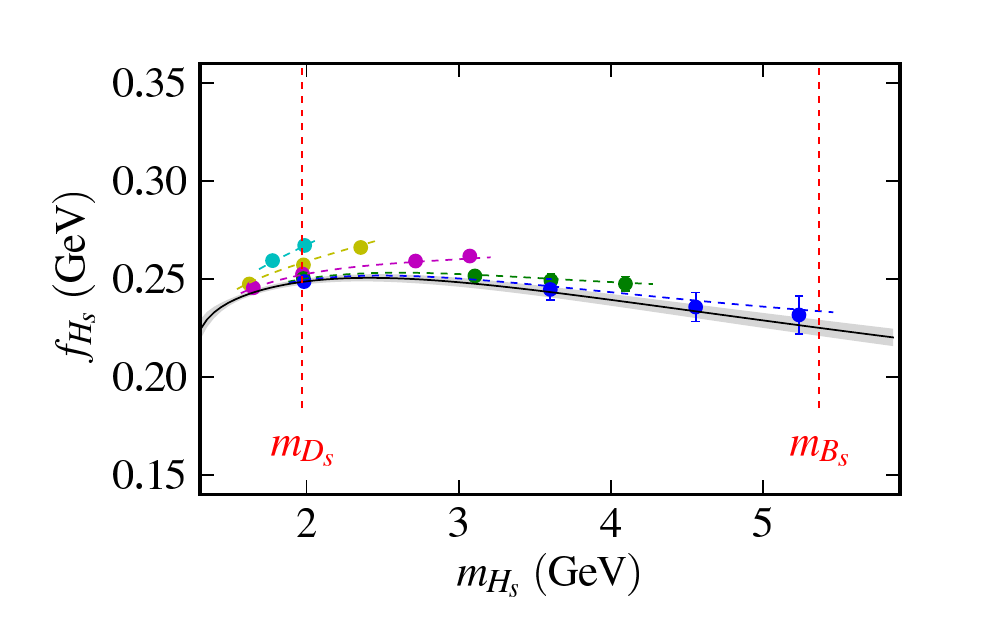}
}
\caption{ Covering the region from charm to bottom: 
a) The ratio of the quark mass in the $\overline{MS}$ scheme at 
a given $\mu$ to one half the pseudoscalar heavyonium mass 
as a function of the pseudoscalar heavyonium mass (as a proxy 
for the heavy quark mass)~\cite{curr}. Note how 
flat the curve for $\mu=m_h$ is. 
b) The Heavy-strange decay constant as a function of heavy-strange 
meson mass (as proxy for heavy quark mass)~\cite{fbs}. Note that the curve 
peaks at the $D_s$. 
}
\label{fig:ctob}
\end{figure}

These results can be 
compared to the HPQCD result of 
$m_b(m_b)^{n_f=5}$ = 4.164(23) GeV~\cite{curr}, 
obtained using current-current correlator methods with 
HISQ quarks on MILC $n_f=2+1$ configurations 
and perturbatively matching to $n_f=5$. 
Here a range of heavy quark masses were explored from $c$ upwards 
at 5 different values of the lattice spacing, allowing the 
physical curve of the ratio of the quark mass to one half of 
the pseudoscalar 
heavyonium mass to be mapped out. This is interesting 
because it allows lattice QCD to `fill in' the values between 
the two results at $c$ and $b$ on which we have experimental 
information.
What we see, reproduced in Fig.~\ref{fig:ctob}a, is that 
$m_{\eta_h}/2m_h(\mu)$ gives a very flat curve with value
falling between 1.2 and 1.1 as $m_h$ is increased when 
$\mu = m_h$. This is a well-defined and accurately quantifiable   
version of the hand-waving statement that "the heavy quark 
mass is roughly half the heavyonium meson mass" and only 
possible from lattice QCD. 

It would be useful to have accurate results for $m_b$ from 
NRQCD, Fermilab and RHQ methods for comparison - these are underway. 
  
\subsection{ Leptonic decays (to $l\nu_l$) }

\begin{figure}[h]
\parbox{0.45\hsize}{
$a)$ \includegraphics[width=0.95\hsize]{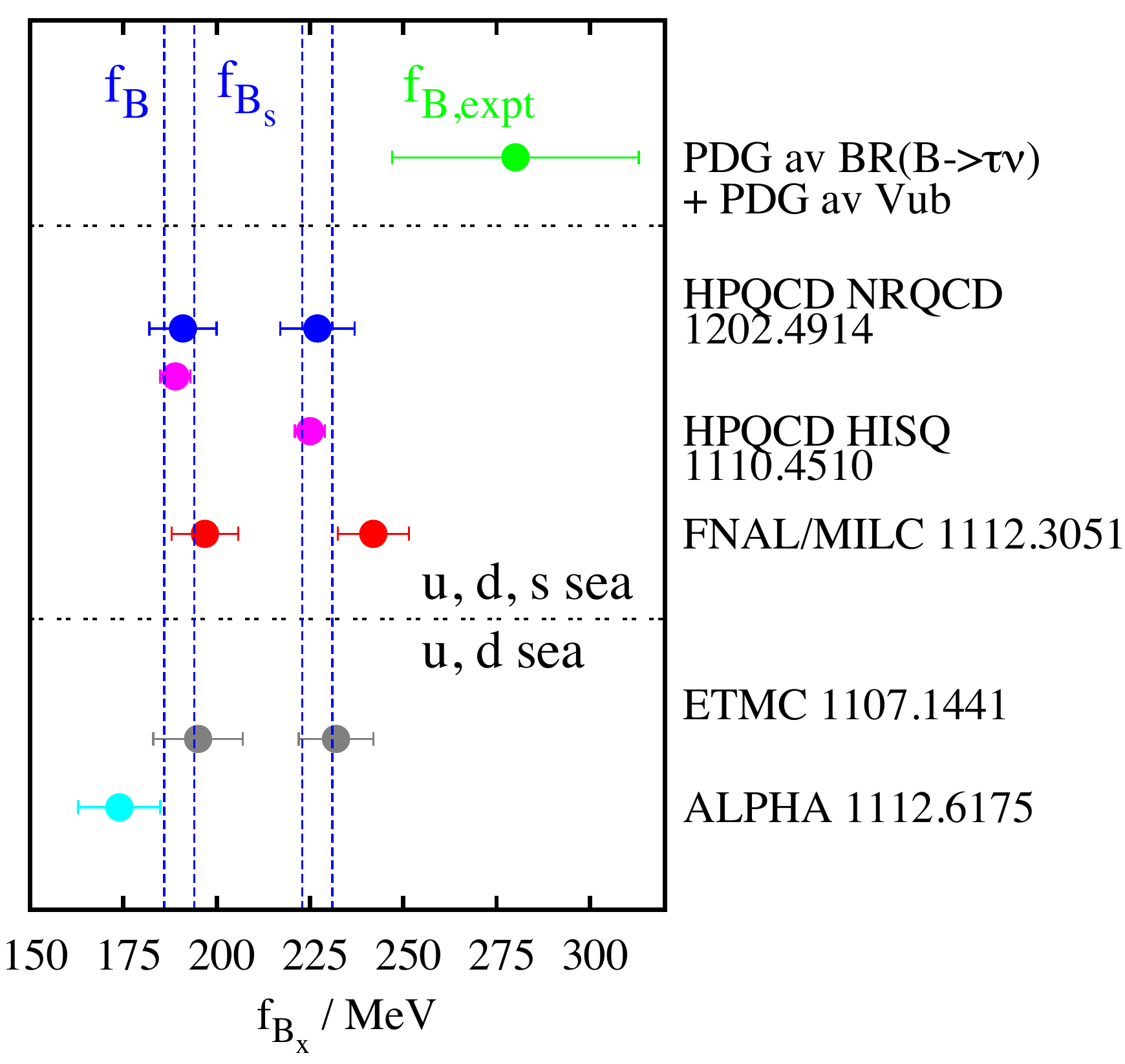}
}
\parbox{0.55\hsize}{
$b)$ \includegraphics[width=0.95\hsize]{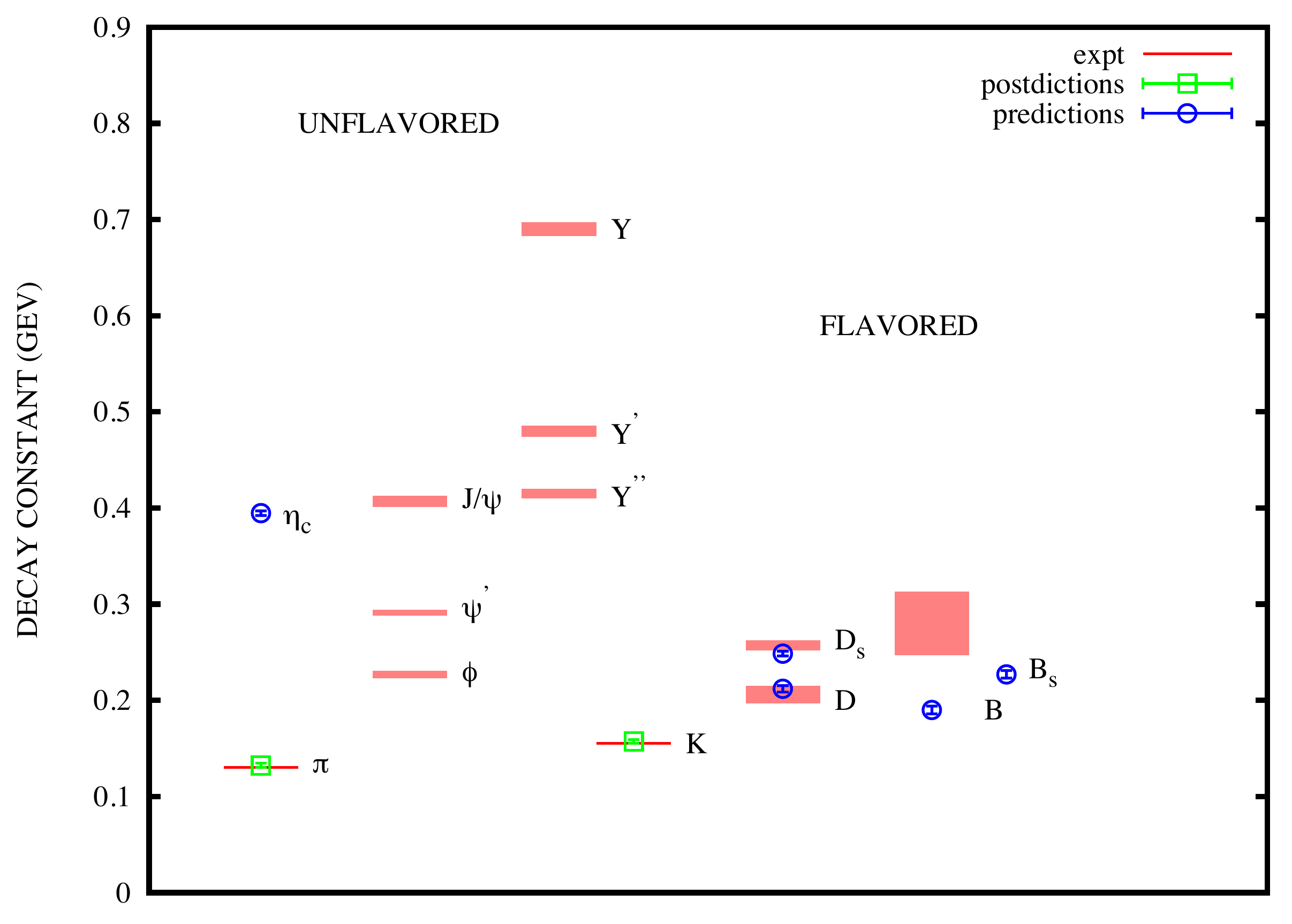}
}
\caption{ 
a) Comparison of lattice QCD results for $f_{B_s}$ and $f_B$ 
and the experimental 
average for $f_B$ from $B \rightarrow \tau\nu$ decay, using $V_{ub}=3.47 \times 10^{-3}$. 
The lattice QCD world averages are $f_B$=190(4) MeV and $f_{B_s}$=227(4) MeV (see text), 
shown with dashed lines. 
b) The `spectrum' 
of decay constants, using $D/D_s/B/B_s$ world averages from here, 
$\pi/K$ from~\cite{oldfds} and $\eta_c$, surprisingly close to $J/\psi$ from experiment, from~\cite{fds}. 
}
\label{fig:fb}
\end{figure}

Direct annihilation to a $W$ is only possible for charged pseudoscalars. 
However it is still possible and useful to calculate the equivalent 
matrix element for the neutral $B_s$ meson in lattice QCD 
and compare its value 
to that for the $B$.
There are several new results this year for $f_B$ and $f_{B_s}$ and 
these are summarised in Fig.~\ref{fig:fb}. 

The two using nonrelativistic methods on the MILC $n_f=2+1$ asqtad 
configurations can be directly compared. These are the HPQCD results 
using NRQCD $b$ quarks (without the further improvements discussed 
above) with HISQ valence light quarks~\cite{junko} and the Fermilab/MILC 
results using Fermilab $b$ quarks and asqtad light quarks~\cite{fnalnewf}. The 
HPQCD NRQCD-HISQ results of $f_{B_s}$ = 227(10) MeV and $f_B$=191(9) MeV 
supersede earlier NRQCD-asqtad calculations~\cite{gamiz} using the same 
methods - 
the error is improved by 50\% because of better statistics and 
smaller discretisation errors but the central values change very 
little. The Fermilab/MILC results of $f_{B_s}$ = 242(10) MeV 
and $f_B$ = 197(9) MeV are calculated on essentially the same set 
of $a=$0.12 and 0.09fm gluon fields and also update earlier 
work. The Fermilab/MILC and HPQCD results agree reasonably well - 
$f_{B_s}$ values differ by 1.5 $\sigma$ allowing for correlated statistical 
errors. In both cases the error budgets are dominated by other 
errors, presumably uncorrelated. For the NRQCD case the problem
 is the unknown $\mathcal{O}(\alpha_s^2)$ 
pieces of the perturbative matching of the axial current to the 
continuum. The error allows for 
a coefficient at $\mathcal{O}(\alpha_s^2)$ of 0.4 (the one-loop 
coefficient is around 0.1), 
uncorrelated between the two lattice spacing 
values since it can vary with $m_ba$ (the $\mathcal{O}(\alpha_s)$ 
coefficient varies from 0.1 to 0.15). For the Fermilab 
case the dominant errors are heavy-quark discretisation and tuning 
and statistics. Their perturbative matching error allows for 
$0.1\alpha_s^2$in the ratio of axial to vector renormalisations, as described earlier for the $D/D_s$ case. 

$f_{B_s}/f_B$ is determined to around 2\% in both cases with values 
1.188(18) for NRQCD-HISQ and 1.229(26) for Fermilab-asqtad. Again this 
represents agreement at the 1.5$\sigma$ level allowing for correlated 
statistical errors. Now statistical errors are a more significant 
part of the total since the renormalisation and heavy-quark discretisation 
effects largely cancel. Neither result for $f_{B_s}/f_B$ is sufficiently 
accurate to distinguish it from $f_{D_s}/f_D$ or $f_K/f_{\pi}$ although 
the statistical and systematic 
correlations in the Fermilab/MILC results between the $B$ and $D$ 
calculations presumably 
mean that the difference between their values of 
1.229(26) for $f_{B_s}/f_B$ and 
1.188(25) for $f_{D_s}/f_D$ is significant. 
The Fermilab/MILC results will improve with values on additional 
finer MILC asqtad ensembles; the HPQCD results will be updated with 
improved NRQCD on the HISQ 2+1+1 configurations, exploring also 
improvements in renormalisation that may be possible with heavy-light 
current-current correlator methods~\cite{jonnalat10}.  

The Alpha collaboration gave an updated result from their HQET 
method for $f_B$ only, obtaining $f_B$=174(11) MeV~\cite{alphalat11}. 
This is done on the CLS $n_f=2$ lattices with 3 values of $a$ from 
0.075 fm to 0.05 fm and $m_{\pi}$ values down to 270 MeV. 
The $b$ quark is included in the static (infinite mass) formalism 
using an action with smeared links. This improves the poor signal/noise 
for static quarks which arises from the unphysical 
energy of the heavyonium state (with no kinetic term) 
that contributes to the noise. 
Matrix elements of operators that give rise to 
$1/m_b$ corrections to the static limit are included after 
a nonperturbative determination of their coefficients. The 
result for $f_B$ is 1.5$\sigma$ lower than those from the 
2+1 flavor nonrelativistic results discussed above. 

Relativistic methods have yielded well-controlled results this year
for the first time and these look encouraging for the future. 
ETMC use HQET to extrapolate up to the $b$ quark from quark masses around 
$2m_c$ using 4 values of $a$ down to 0.05fm on $n_f=2$ 
lattices~\cite{tmfb}. They include a 
static result to bound the upper limit of 
the extrapolation. They average over two different methods for 
$f_{B_s}$ that give results 225(8) MeV and 238(10) MeV, possibly 
worryingly far apart given that the (correlated) statistical errors 
dominate. The final result for $f_{B_s}$ is 232(10) MeV and a separate 
determination of the ratio $f_{B_s}/f_B$ is used to obtain $f_B$ = 195(12) MeV. 
HPQCD use HISQ quarks~\cite{fbs} with masses from slightly below $m_c$ up to 
close to $m_b$ to map out the heavy-strange decay constant as 
a function of the heavy-strange meson mass (as a physical proxy 
for the heavy quark mass). 
They use 5 values of $a$ down to 0.045 fm on $n_f=2+1$ 
configurations and limit 
$m_ba$ to values below 0.85. The method relies on the 
finding that $f_{D_s}$ is very insensitive to sea quark 
masses~\cite{fds} to avoid an extensive study as a function of 
sea light quark mass. The extrapolation to the $b$ uses (and tests) 
HQET formulae and obtains $f_{B_s}$ = 225(4) MeV, significantly 
more accurate than previous results. Surprisingly 
it provides the first really solid 
demonstration (although most earlier results have indicated this) 
that $f_{B_s} < f_{D_s}$ with the 
ratio $f_{B_s}/f_{D_s}$ = 0.906(14). 
In fact $f_{D_s}$ seems 
to be close to the maximum of the heavy-strange decay constant
curve (see Fig.~\ref{fig:ctob}b), which shows a rapid rise from lower 
masses up to $f_{D_s}$ and then a rather slow fall down to 
$f_{B_s}$, much slower than that of leading-order HQET. 
Since the ratios of 
heavy-strange and heavy-light decay constants for $B$ and $D$ differ very 
little, note that it is also true that $f_B < f_D$. 

To obtain a world-average value of $f_{B_s}$ I take an 
error-weighted average of the three $n_f=2+1$ results, allowing 
100\% correlation between the statistical errors of the 
Fermilab and NRQCD results.
The HPQCD-HISQ result I treat 
as independent since it relies on results on finer lattices. 
The world-average is then $f_{B_s}$ = 227(4) MeV. 
For a world-average value of $f_B$ I average the Fermilab result 
with the HPQCD value of $f_B$ = 189(4) MeV~\cite{junko} 
that uses the HISQ value 
for $f_{B_s}$ and the NRQCD result for $f_{B_s}/f_B$. This 
gives 190(4) MeV. Both of these new world-average values are lower than 
last year by 1-2$\sigma$ (see for example~\cite{latav}) 
and the error has improved by a factor 
of 3. Note that the lattice QCD result for $f_B$ 
is now clearly below 200 MeV. 

Experimental observation of $B$ leptonic decay 
to $l\overline{\nu}_l$ is extremely 
difficult and the errors on the results consequently rather 
large. Since the rate is proportional to $m_l^2$ (eq.~\ref{eq:leptrate}) 
it is not surprising that the mode that has been seen is 
$B \rightarrow \tau \nu$. The PDG~\cite{pdg} give the average of 
the BaBar and Belle branching fractions as $1.65(34) \times 10^{-4}$. 
Putting in the kinematic factors and $B$ lifetime, 
this corresponds to:   
\begin{equation}
f_B|V_{ub}| = 0.97(10) \,\,  \mathrm{MeV} .
\label{eq:exptfb}
\end{equation}
$V_{ub}$ is also rather uncertain with sizeable differences 
between that extracted from inclusive and exclusive (using 
lattice QCD) $B$ semileptonic 
decay modes. The PDG~\cite{pdg} give a result for $V_{ub}$ from requiring 
unitarity of the CKM matrix as $3.47(16) \times 10^{-3}$. This 
central value is fairly close to that from 
exclusive decays - the inclusive result 
is higher at $4.3(4) \times 10^{-3}$. Combining the unitarity $V_{ub}$ 
with eq.~\ref{eq:exptfb} gives $f_{B, \mathrm{expt}} =$ 280(33) MeV which
is over 2.5$\sigma$ higher than the lattice QCD result above. This is 
a cause of `tension' in the CKM picture~\cite{lunghi} which will need improved 
experimental results to resolve, although understanding the 
ambiguity in $V_{ub}$ would also help. From eq.~\ref{eq:exptfb} and the 
lattice average for $f_B$ above, we can 
derive $V_{ub} = 5.1(5) \times 10^{-3}$, not that much worse than other 
direct determinations. 

Improvements to lattice QCD $B/B_s$ decay constants will continue - 
it seems likely that relativistic methods will provide the best way 
forward for the absolute value of $f_{B_s}$. 
The ratio of $f_{B_s}/f_B$ may be better calculated with 
nonrelativistic methods and can certainly be obtained 
to 1\% errors working closer to the chiral limit. However, it 
should be remembered that the ratio that 
is really required for combination with experiments on $B_s/B_d$ mixing 
is the ratio of 4-quark 
operator matrix elements and this is harder~\cite{gamiz}.  

\subsection{$B_{(s)} \rightarrow \mu^+\mu^-$} 
An important leptonic mode for neutral $B$ mesons is 
decay to $\mu^+\mu^-$. In the Standard Model this proceeds via 
$Z^0$ penguin and box diagrams~\cite{buras1} and so is expected to 
be sensitive to new physics. The effective weak Hamiltonian that 
gives rise to the decay for $B_s$ is then 
\begin{equation}
H_{\mathrm{eff}} = -\frac{G_F}{\sqrt{2}}\frac{\alpha}{2\pi\sin^2\theta_W}V_{tb}^*V_{ts}Y(x_t)(\overline{b}s)_{V-A}(\overline{l}l)_{V-A} + h.c.  
\end{equation} 
where $Y$ is a known function of the mass of the top quark 
and $x_t = m_t^2/m_W^2$. The 4-fermion operator 
here has 2 quarks and 2 leptons and so, as far as QCD is concerned, 
it looks like an operator for quark-antiquark annihilation. The 
matrix element is then proportional to the decay constant $f_{B_s}$. 
The rate, however, depends on CKM elements $V_{tb}^*V_{ts}$ which 
are derived from the $B_s$ mixing rate ($\Delta M_s$) along with a lattice 
QCD calculation of the matrix element of the 4-quark operator that 
corresponds to the box diagram for that process, and is parameterised 
by $f_{B_s}^2B_{B_s}$. $B_{B_s}$ is known as the `bag parameter'. 
Buras pointed out~\cite{buras2} that in fact the best 
way to determine the rate for $B_s \rightarrow l^+l^-$ was to take 
a ratio to the mixing rate. Then the CKM elements cancel and so does 
$f_{B_s}$ and the SM rate for $B_s \rightarrow \mu^+\mu^-$ is 
proportional to the bag parameter, $B_{B_s}$.  A determination of 
this {\it requires} lattice QCD (despite the confusion over this in the 
literature). 

Only one such lattice QCD calculation including sea quarks has so far been 
done~\cite{gamiz}, by the HPQCD collaboration, as 
part of a calculation to obtain 
the mixing matrix elements for $\Delta M_s$. 
This yields the current best estimate of the $B_s$ bag 
parameter, $\hat{B}_{B_s}$ = 1.33(6) giving a SM branching 
fraction for $B_s \rightarrow \mu^+\mu^-$ of $3.19(19) \times 10^{-9}$. 
The calculation used NRQCD $b$ quarks and asqtad light quarks and perturbative 
renormalisation of the 4-quark operators on MILC $n_f=2+1$ asqtad 
configurations. A significant part of the error was from 
statistics and will be reduced in new calculations underway 
on the new $n_f=2+1+1$ HISQ configurations. 

The observation of 
$B_s \rightarrow \mu^+\mu^-$  is a key aim of the LHC experiments. 
This year~\cite{moriond} LHCb and CMS hav improved their limits 
on the branching fraction 
to within a factor of 3 of the Standard Model rate from lattice QCD. 
In the absence of BSM physics, 
LHC expects to see the process in 2012, and then the comparison 
of theory and experiment will become more critical, and improved 
errors on the lattice QCD side may be very important. 

\subsection{Other decay modes and mixing}
\label{sec:bother}

The Fermilab 
Lattice/MILC collaborations discussed their progress on the 
calculation of 4-quark operator matrix elements for $B_s$ 
and $B_d$ mixing, using Fermilab 
$b$ quarks and studying the complete set of 5 $\Delta B=2$ 
operators~\cite{bouchard}.   
There was also an update on $B \rightarrow K l^+l^-$ form factors 
which may be sensitive to new physics~\cite{zhou}. 

The calculation of the ratio of scalar form factors for $B_s \rightarrow D_s$ 
and $B \rightarrow D$ turns out to be useful for the experimental 
normalisation of the $B_s \rightarrow \mu^+\mu^-$ mode~\cite{fleischer}. 
QCD sum rule calculations~\cite{blasi} expect 
deviations from 1 in this ratio to be 
related, and of similar size, to the 20\% $u/d-s$ effects 
seen in decay constants. 
Instead, as discussed in 
section~\ref{sec:csemil}, very little spectator quark mass 
dependence is seen in heavy form factors~\cite{jonna} 
to a high level of accuracy. 
Fermilab/MILC have now done an explicit calculation of these $B$ and $B_s$ 
semileptonic modes~\cite{fnalbd} and indeed 
find no significant deviation from 
1: $f_0^{B_s}(m_{\pi}^2)/f_0^{B_d}(m_K^2)$ = 1.046(46). 

\begin{figure}[h]
\parbox{0.55\hsize}{
$a)$ \includegraphics[width=0.95\hsize]{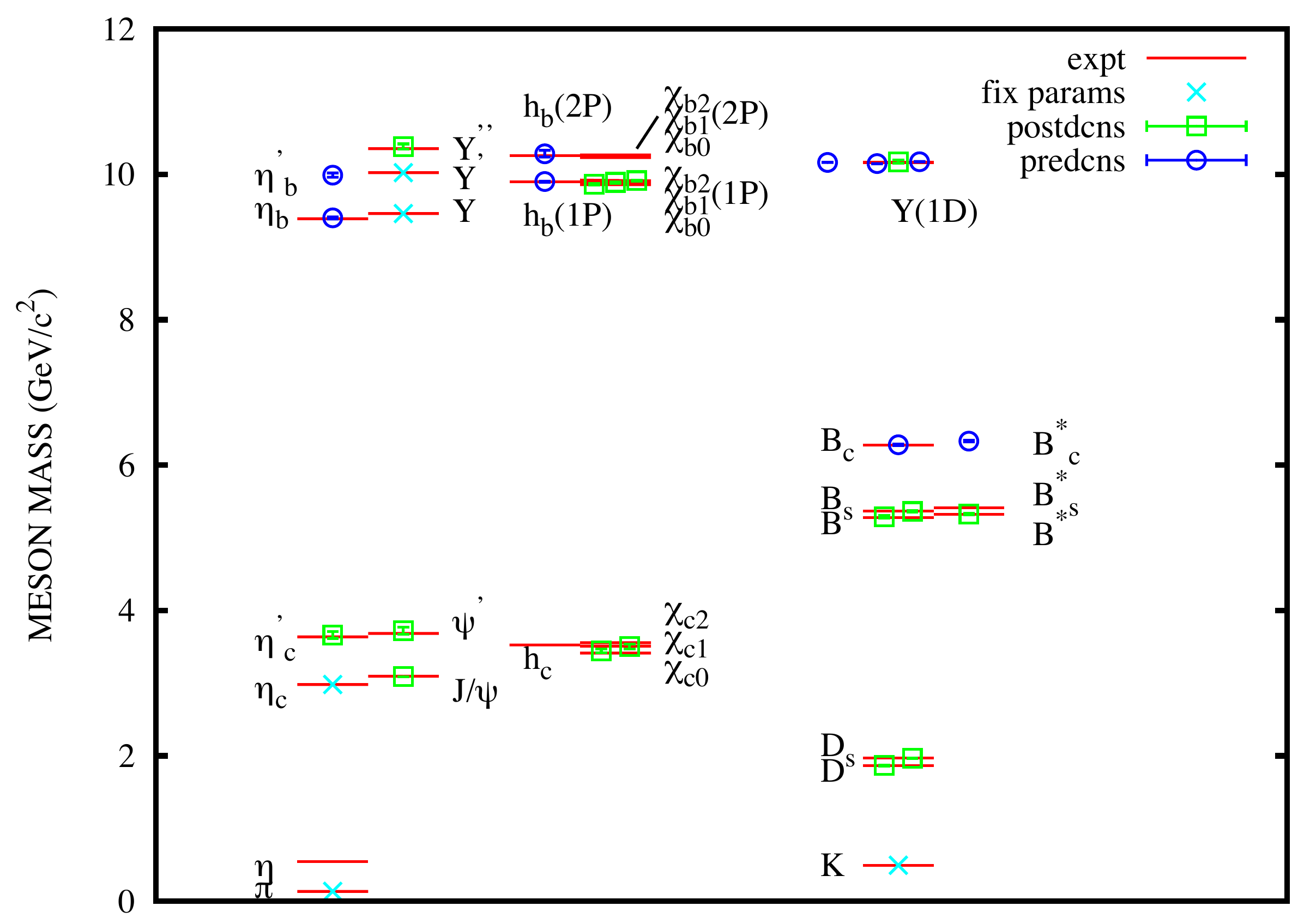}
}
\parbox{0.45\hsize}{
$b)$\includegraphics[width=0.95\hsize]{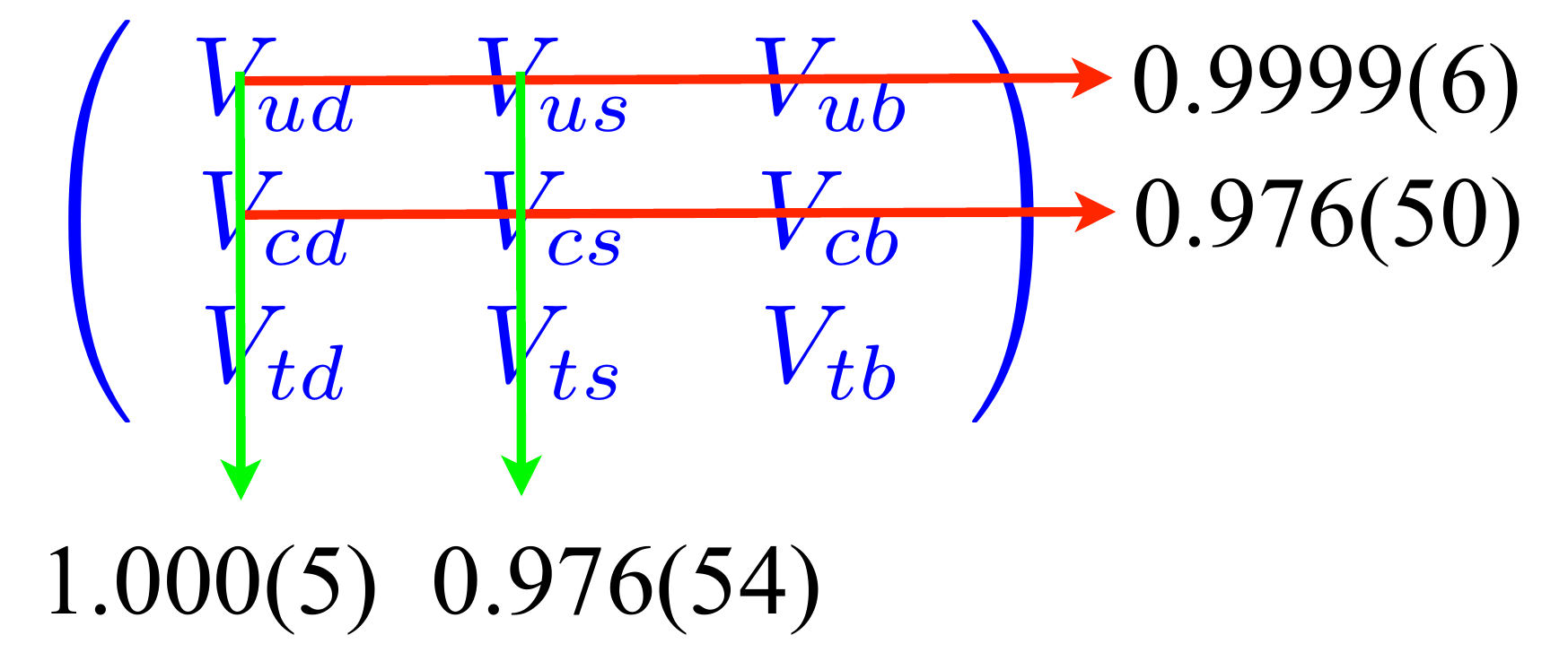}
}
\caption{ a) The spectrum of gold-plated mesons from 
lattice QCD, updating~\cite{gregory2} to include the prediction 
of $D$-wave Upsilon states~\cite{daldrop}. 
b) Tests of unitarity of rows and columns of 
the CKM matrix now 
possible using lattice QCD results, except for $V_{ud}$ from nuclear 
$\beta$ decay~\cite{pdg}. 
I have taken direct 
determinations of $V_{us}$, $V_{cd}$ and $V_{cs}$ from semileptonic modes~\cite{wittig, na1, na2}, 
although there is no particular reason to prefer those over leptonic modes.  
$V_{ts}$ and $V_{td}$ come from lattice QCD $B_s/B_d$ mixing calculations~\cite{pdg, gamiz}
and $V_{cb}$ from $B \rightarrow D^*$ decays ($39.7(1.0) \times 10^{-3}$~\cite{fnalbdstar}) 
but these are too 
small to have much impact on row/column unitarity. 
Unitarity triangle tests, from orthogonality 
of columns 1 and 3, are discussed in~\cite{lunghi}.   
}
\label{fig:goldspectrum}
\end{figure}

\section{Conclusions}
\label{sec:conclusions}
Key results can be summarised in three plots, all showing significant 
progress since last year. Fig.~\ref{fig:goldspectrum}a 
updates the spectrum of gold-plated mesons, which includes many
$c$ and $b$ states. Fig.~\ref{fig:fb}b is a similar 
plot laying out the picture for decay constants, now rather
impressive for pseudoscalars.  
It is clear that similarly accurate results on electromagnetic annihilation 
of vector mesons would 
significantly enhance our confidence in lattice QCD results and 
provide excellent tests of QCD itself. 
Fig.~\ref{fig:goldspectrum}b shows the tests of unitarity of the first 
and second rows and columns of the CKM matrix 
that are now possible at the 5\% level using lattice QCD and experiment. 

\begin{flushleft}
{\bf Acknowledgements} I am grateful to very many people for 
useful discussions, some of them referenced through their talks below.   
Thanks also to the organisers for an excellent meeting. My work is funded by STFC and the Royal Society. 
\end{flushleft}

\end{document}